\documentclass[12pt]{article}
\usepackage{amssymb}
\usepackage{graphicx}
\usepackage{float}
\usepackage{amsmath}
\usepackage{epsfig}
\usepackage{latexsym}
\usepackage{bm}
\usepackage{natbib}
\usepackage{amssymb}
\usepackage{color}
\usepackage{amsmath,bm,epsfig,subfigure}
\usepackage{epstopdf}
\begin{document}
%newcommand for bold italic
\newcommand{\bfit}[1]{\textit{\textbf{#1}}}
\makeatletter
\renewcommand{\fnum@figure}{Fig. \thefigure}
\makeatother

\title{The Response of Conductive-Fiber Reinforced Composites to Electric Field }
   \author {
  {\Large  Nathan Perchikov and Jacob Aboudi$^*$ }      \\
   School of Mechanical Engineering \\
   Faculty of Engineering           \\
   Tel Aviv University              \\
   Ramat Aviv 69978, Israel         \\
   $^*$ Corresponding author: E-mail: aboudi@eng.tau.ac.il,  \\
   \ tel: +972-3-6408131, \ fax: +972-3-6407617 \\
  }
\date{  }

\maketitle

\newpage
\begin{abstract}

An analytical procedure, which couples electric, magnetic, thermal and mechanical effects is presented for the prediction of the response of unidirectional fiber-reinforced
composites that are subjected to electric field, applied in the fibers' direction.
It is assumed that at least one of the phases of the composite (e.g., the fibers) is electrically conductive, and that all phases are thermally conductive.
The composite is assumed to occupy a finite, symmetric domain which is discretized into a double array of subcells.
The governing equations, with the interfacial and boundary conditions, are satisfied in the integral sense.
The externally applied field generates electric current, which induces a magnetic field as well as temperature increase.
The mechanical deformation of the composite results from the combined effect of the ponderomotive force, which is created by the magnetic field,
and the temperature distribution within the constituents.
The purpose of the present paper is three-fold.
(a) To perform quantitative analysis of the model for the ponderoromotive force in deformable media.
(b) To present computational strong-form treatment of the magneto-mechanical boundary-value problem in a composite.
(c) To suggest a computational apparatus for deriving the response of a sensor/actuator excited by an applied electric field or electric field gradient.    
Application is given, presenting the magnetic, thermal and mechanical field distributions, as well as the macroscopic (global) response of a composite, which consists,
for simplicity, of two iron fibers embedded in an epoxy matrix.

\


\end{abstract}

\newpage
\section{Introduction}
\label{Sect1}

The micromechanical response of composite materials is usually investigated when the application of mechanical or thermo-mechanical loading is considered.
Additional macroscopic physical effects, such as electric and magnetic field distribution, are commonly treated for materials with constitutive coupling,
such as in the case of piezoelectric and electrostrictive or magnetostrictive materials. Fewer work deals with the response to electric loading of electrically
conductive composites, as, for example, in the case of a composite in which one of the phases is electrically conductive (e.g., carbon fibers).
In such a case, as a result of the application of electric field, electric current flows in the conductive phase, inducing magnetic field distribution, as well as temperature increase.
The magnetic field gives rise to the so-called ponderomotive, or generalized Lorentz force, which, along with the corresponding temperature distribution, generates mechanical deformation.

The main challenge in addressing the aforementioned type of problems is the proper description of Maxwell's equations and the associated Maxwell tensor, or alternatively,
the ponderomotive force and moment couple density, for the case of deforming continua.
Fundamental investigations can be found in the work of \cite{Tiersten}, \cite{Pao} and in the recent reviews of \cite{Hutter} and \cite{Santapuri2013}.

Beyond the establishing of the constitutive and governing equations of the aforementioned 'multi-physics' problem,
there arise mathematical issues pertinent to the specifics of the field form of Maxwell's equations.
In the case of slow electromagnetic dynamics, the main issue is related to the fact that the electric and magnetic fields each have to satisfy multiple differential equations.
In the context of numerical solution, this difficulty is usually treated by introducing auxiliary gauges, e.g. the Coulomb gauge,
see \cite{Zienkiewicz}, \cite{Simkin} and \cite{BC}, for example. Another approach is taken in \cite{TK},
where a hyperbolic system of evolution equations is derived with no auxiliary differential equations to impose.

Computationally, the coupled multiphysics response of composite materials for general three-dimensional domains is normally derived using
the finite element formalism, see \cite{Sridhar} for a recent example.
In the case of plates, the coupled problem is solved by integrating the plate equations using finite difference discretization,
in conjunction with the Love-Kirchhoff, as in  \cite{ZS07}, \cite{ZS11}, \cite{BZ14} or the von K\'arm\'an approximation, as in \cite{Librescu2003}.

In \cite{Librescu2003}, the dynamic response of a plate reinforced by electrically conducting fibers is considered.
As a result of the nonlinear dependence of the mechanics on the magnetic field, linearization is performed, based on the assumption of small disturbances. 
In \cite{BZ14} on the other hand, a similar setting is assumed, also for plate dynamics, however without the small disturbances assumption.
A different model for the ponderomotive force is used, and much as in \cite{Librescu2003}, uniform material, rather than a composite, is considered. The solution is obtained by a finite difference method. 

The purpose of the present article is three-fold. 
First, to perform quantitative investigation of the effect of the correction for the ponderomotive force, relatively to the standard expression
of the Lorentz force, which is established for the particle-in-vacuum case.
To this end we follow \cite{Santapuri2013} in choosing the Maxwell-Minkowski model, reduced for the linear quasistatic regime.
The framework for the examination of the aforementioned effect is that of unidirectional deformation of a two-constituent composite specimen.
The second perspective in which the present work is performed is computational.
To this end, the Higher Order Theory (HOT), as in \cite{AAB}, which has been employed for the analysis of functionally graded materials, is generalized to account
for electric-current-induced mechanical deformations. In the framework of the HOT, the governing equations together with the interfacial and boundary conditions are imposed in the average (integral) sense.
It should be noted that since the HOT is a strong-form approach to the solution of boundary-value problems,
the present work opts to construct a strong-form computational apparatus for the analysis of the multi-physics response of composites in the linear quasistatic regime.
In the third perspective, the present paper in effect proposes a sensor/actuator with the accompanying computational apparatus,
which will have the effect of nonlinear response to either electric field or electric field gradient, in line with the view presented in \cite{Santapuri2015}.
The details of these perspectives are further discussed in the concluding section. 

To elaborate on the third perspective, the present article offers a coupled electro-magneto-thermo-mechanical analysis for the prediction of the mechanical deformation of partially conductive
composites, caused by the application of electric field.
The engineering motivation for this analysis can be illustrated by the following discussion.
Consider a setting (e.g. a large capacitor), in which a uniform electric field exists. It is then assumed that a local effect (perturbation -- a small relative change in the field but with a high local gradient, as created by a small conducting object) has been introduced, as a result of which the electric field loses its uniformity.  
The present analysis shows that a device (sensor) in the form of a ring that consists of, say, two conductive wires embedded in a polymeric (insulation) matrix, will reveal the effect by showing the induced mechanical deformation.
This deformation, caused jointly by the Joule heating and the ponderomotive force,
can be calibrated to estimate the intensity of the perturbation-induced local electric field gradient.
Such a (partially conducting) ring would be advantageous over, say, a piezoelectric sensor
in that it would sense the finite electric field gradient rather than the nearly uniform
local field itself. High accuracy of the device-calibration would be obtained due to the offered analysis,
owing to it being based on strong-form solution
of the equations (which is possible for linear constitutive relations).  
It should be noted that a sensor in the form of a ring is a three-dimensional object,
and a 3D imperfectly conducting ring in a non-uniform electric field develops a current, see \cite{Assis}.
It is also shown in that article that the expressions for the 3D corrections to the electric field
in the ring become negligible for a slender wire.
Consequently, in the limiting case of a slender ring-wire the applicability of the present 2D analysis is justified.

In the present article, the analysis of mechanical deformation generated by the application of externally applied electric field on electrically and thermally conducting
fiber-reinforced composites is considered.
The composite material is assumed to consist of continuous fibers that are distributed in the matrix, forming a symmetric array.
The electric field is applied in the fibers' direction, and the boundaries of the composite are assumed to be traction-free and thermally convective.
The composite region is divided into a double array of subcells.
The magnetic field induced by the applied electric field is derived, and the resulting ponderomotive force is established.
Next, the equations that govern the thermal problem related to the heat generated by the electric current are established
and solved numerically, to obtain the temperature field distribution and its gradient.
Finally, the mechanical deformations of the composite phases are obtained from the solution of a system of sparse algebraic equations, following the HOT approach.
The presented derivation is illustrated for the case in which uniform (throughout the composite domain) electric field is applied to a thermally conductive polymeric matrix (epoxy) reinforced
by two iron fibers, which are both electrically and thermally conductive.
The solution demonstrates the distributions of the magnetic, thermal and mechanical fields
created by the interaction between the two fibers.
It should be emphasized, however, that the present theory can be generalized to establish the response of any number of fibers of identical or different properties and cross-section.
Such generalization would affect only the computational complexity caused by the increase in the number of subcells.

\begin{figure}[H]
\begin{center}
{\includegraphics[scale = 0.7]{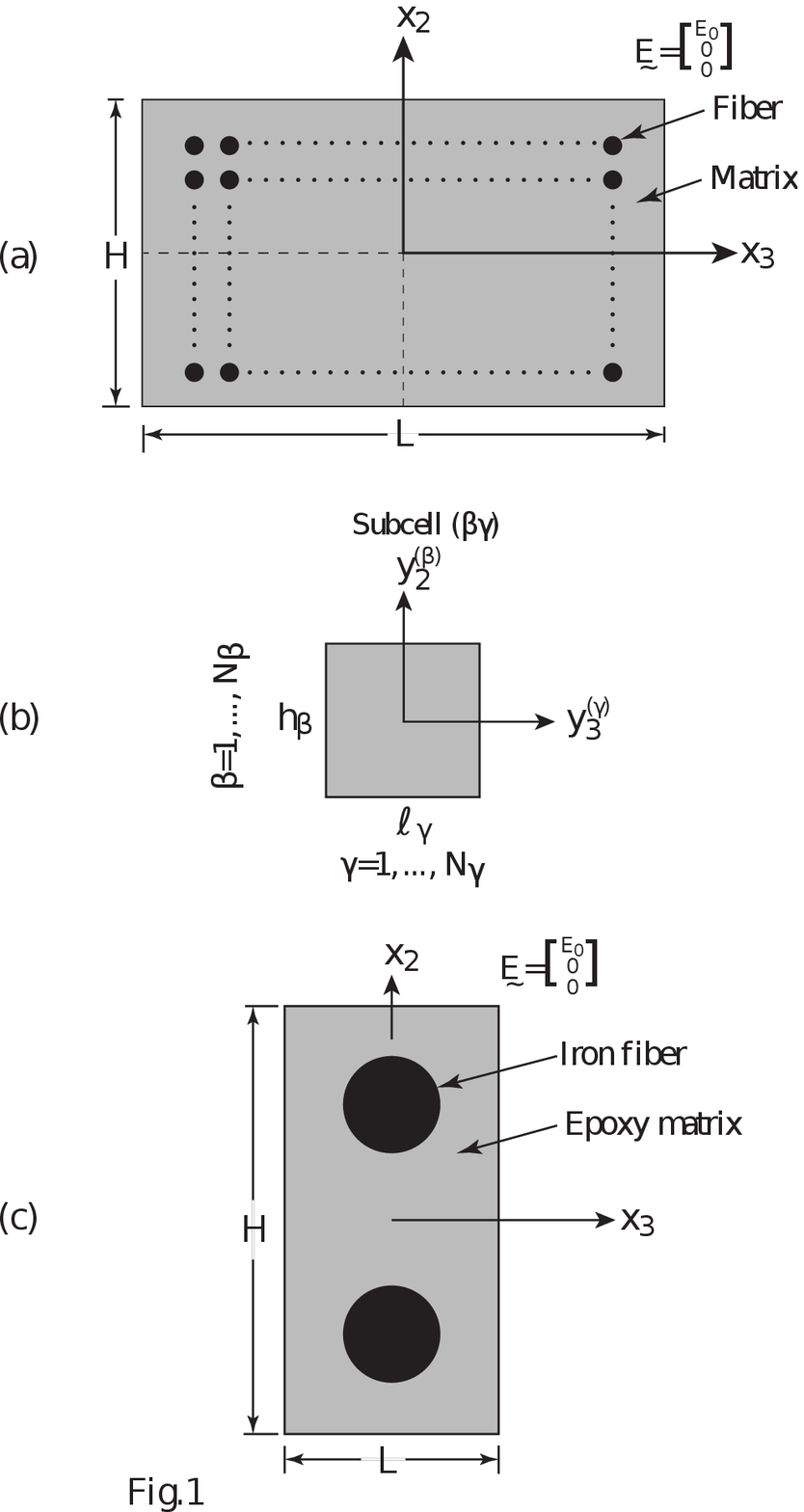}}
\end{center}
\caption{\small (a) A fiber-reinforced composite, described with respect to global coordinates $(x_1, x_2, x_3)$, in which the continuous fibers, oriented in the $x_1$-direction,
                      are symmetrically distributed. The composite is subjected to electric field $E_0$ in the $x_1$-direction. 
                  (b) A typical subcell $(\beta \gamma)$, in which local coordinates $(y^{(\beta)}_2, y^{(\gamma)}_3)$ have been 
                      introduced whose origin is located at the subcell center.
                  (c) Fiber-reinforced iron/epoxy composite consisting of two fibers.}
\label{Fig1}
\end{figure}

The structure of this paper is as follows. In Section 2, the governing equations are presented.
In Sections \ref{Sect3}, \ref{Sect4}, \ref{Sect5} and \ref{Sect6} the electric, magnetic, thermal and mechanical fields, respectively, are considered,
and the procedures for their computation are derived. In Section \ref{Sect7}, the application of the derived computational method is demonstrated.
The obtained numerical results are discussed in Section \ref{Sect8}. The conclusions section, Section \ref{Sect9}, provides overall discussion of the derived theory.

\section{Governing Equations}
\label{Sect2}

Consider a unidirectional fiber-reinforced composite, in which the fibers are as shown in Fig. \ref{Fig1}1(a). 
The composite occupies the region $-H/2 \le x_2 \le H/2$, $-L/2 \le x_3 \le L/2$, and it is assumed that the domain is symmetrically occupied with respect to the axes. 
The fibers are assumed to be electrically conductive. 
A uniform electric field, namely, ${\bm E_0} = \left\{E_0, 0, 0 \right\}^\top$ is acting at the boundary in the out-of-plane x$_1$-direction (fibers' direction), formally 'at $\pm \infty$'.
As a result of the existence of the conductive phase, electric current is generated and magnetic field is induced.
The corresponding temperature increase and ponderomotive force produce mechanical deformation of the composite.  

The governing equations are given by Maxwell's equations, along with a model for the ponderomotive force, energy balance and mechanical equilibrium (momentum balance), as follows.
Starting with Faraday's law for the static regime and Gauss's law for the case of  uncharged material:
\begin{eqnarray} \label{E0}
\nabla \cdot {\bm D} = {0}, \ \ \ \ \nabla \times {\bm E} = 0 
\end{eqnarray}
where $\bm{D}$ is electric displacement vector and $ {\bm E}$ is the emergent electric field intensity in the composite, and following by magnetic flux conservation and Amp\`ere's law:
\begin{eqnarray} \label{E1}
\nabla \cdot {\bm B} = {0}, \ \ \ \ \nabla \times {\bm H} = {\bm J} 
\end{eqnarray}
where ${\bm B}$, ${\bm H}$ and ${\bm J}$ are the magnetic flux density, magnetic field intensity and electric current density, respectively.
The energy balance in the linear elastic quasistatic limit reads:
\begin{eqnarray} \label{E2}
\nabla \cdot {\bm q} = {\bm J} \cdot {\bm E}
\end{eqnarray}
where ${\bm q}$ is the heat flux.
Next, the mechanical equilibrium equation reads:
\begin{eqnarray} \label{E3}
\nabla \cdot {\bm \Sigma} + {\bm F} = {\bm 0} 
\end{eqnarray}
where ${\bm \Sigma}$ and ${\bm F}$ are the Cauchy stress tensor and body force-per-unit-volume vector, respectively.

The corresponding constitutive equations are as follows.
The magnetization law is assumed to be linear, even though in the application section a ferromagnetic material is chosen
in order to enhance the effect of the ponderomotive force on the deformation.
The resolution of this seeming contradiction is simply the assumption that the magnetic field intensity remains small,
corresponding to the assumption of an applied weak electric field excitation at the boundary.
In addition, quasistatic increase of the electric field at the boundary is assumed, which corresponds to linear response with no appearance of magnetic hysteresis: 
\begin{eqnarray} \label{E4}
{\bm B} = \mu_0 \mu_r {\bm H}
\end{eqnarray}
where $\mu_0$ is the free space permeability and $\mu_r$ is the constant relative material permeability.
Ohm's law in the vanishing material velocity limit and the electric polarization equation in the weak field approximation in terms of the electric displacement vector reads:
\begin{eqnarray} \label{E5}
{\bm J} = \sigma {\bm E} \ \ , \ \ {\bm D} = \epsilon_0 \epsilon_r {\bm E}
\end{eqnarray}
where $\sigma$ is the constant coefficient of electric conductivity and $\epsilon_r$ is the
relative constant electric permittivity coefficient of the material, and where $\epsilon_0$ is vacuum permittivity.
Fourier's law of heat conduction for thermally isotropic materials in the quasistatic limit and small temperature gradient/change assumption: 
\begin{eqnarray} \label{E6}
{\bm q} = - \kappa \nabla \theta
\end{eqnarray}
where $\kappa$ is the constant scalar heat conductivity of the solid and $\theta$ is the temperature field measured as deviation from a uniform reference temperature.
Finally, the Cauchy stress tensor ${\bm \Sigma}$ is related to the small strain tensor ${\bm \epsilon}$ and the temperature deviation $\theta$ as follows:
\begin{eqnarray} \label{E7}
{\bm \Sigma} = {\bm C} {\bm \epsilon} - {\bm \Gamma} \theta
\end{eqnarray}
where ${\bm C}$ and ${\bm \Gamma}$ are the constant material stiffness fourth-order and thermal stress second-order tensors, respectively.
The body force-per-unit-volume vector in Eq. (\ref{E3}) is the ponderomotive force,
representing the macroscopic effect of long-range magnetic dipole interactions.
Here the Maxwell-Minkowski model is taken, following \cite{Santapuri2013}, which in the linear quasistatic regime takes the following form:
\begin{eqnarray} \label{E8}
 {\bm F} = {\bm J} \times {\bm B} + \mu_0 (\mu_r -1) (\nabla {\bm H}^\top ) {\bm H} +(\nabla {\bm{E}}^\top) {\bm{P}}
\end{eqnarray}
where $\bm{P}$ is electric polarization. The distributed couple $\bm{L}$ appearing in the Maxwell-Minkowski model in the dynamic case vanishes for
the quasistatic regime assumed herein, rendering the Cauchy stress tensor symmetric (see \cite{Santapuri2016}). 

In the framework of the present two-dimensional analysis, the external electric field is assumed to be tangential to the global boundaries. Consequently,
the electric boundary condition is that of tangential field continuity.
The magnetic boundary conditions are those of tangential field intensity and normal flux density continuity at the external boundaries.
The mechanical boundary conditions are given by zero tractions imposed at $x_2 = \pm H/2$ and $x_3 = \pm L/2$, together with a suitable displacement pinning to prevent singularity.
The thermal boundary conditions are assumed to be convective.
The interfacial conditions require the continuity of the tractions and normal components of magnetic flux density and heat flux.
In addition, the temperature, displacements and tangential electric and magnetic field intensities must be continuous at the interfaces between the phases.
According to the present method of solution, the composite region $-H/2 \le x_2 \le H/2$, $-L/2 \le x_3 \le L/2$ is divided into
$N_{\beta}$ and $N_{\gamma}$ subcells in the x$_2$ and x$_3$ directions, respectively.
Each subcell is labeled by the indices $(\beta, \gamma)$ with $\beta=1,...,N_{\beta}$ and $\gamma=1,...,N_{\gamma}$.
The dimensions of subcell $(\beta \gamma)$ are denoted by $h_{\beta}$ and $l_{\gamma}$, and 
local coordinates $(y^{(\beta)}_2, y^{(\gamma)}_3)$ are introduced within each subcell whose origin is located at its center, see Fig. \ref{Fig1}(b).  

It should be mentioned that the composite materials which are dealt with in the present article
are not assumed to possess periodic microstructure. Thus a homogenization technique and its associated periodic
boundary conditions are not employed in the proposed analyses. 

\section{The Electric Field}
\label{Sect3}

The electric problem consists of the combination of Eqs. (\ref{E0}) and (\ref{E5}b), and suitable boundary conditions.
The approach followed in the present paper is to divide the spatial domain occupied by the composite to subcells containing each only one material.

In every such subcell, the material properties are uniform and there is no discontinuity in the fields.

Consequently, the differential form of Maxwell's equations is valid. 

The corresponding boundary conditions are then continuity of the fields along the interfaces between the subcells.
In the present electric problem, a possible solution of the aforementioned equations is a field $\bm{E}$ uniform across a given subcell.

Since the global boundary conditions require the electric field to point in the $x_1$ direction only,
and since the interfaces of a subcell in the $x_2-x_3$ plane are oriented perpendicular to $x_1$,
clearly if one assumes the electric field for every value of $x_1$ in the composite to point in the $x_1$ direction, just as the applied field,
then the continuity conditions of the tangential components of the electric field are satisfied. 

It is known that tangential continuity in electric problems should be enforced on the electric field.
For materials with different relative permittivities, electric displacement continuity along the interface may result in an electric field that is non-uniform across material boundaries.

However, since a uniform electric field pointing in the $x_1$ direction satisfies the differential equations in both the fibers and the matrix,
and the only interface continuity condition is that this electric field should be continuous across the boundaries as well,
evidently the emergent electric field in the composite should be just as if the composite were uniform monolithic material:
a uniform field in the $x_1$ direction, insensitive to material boundaries, namely: 

\begin{eqnarray} \label{E9}
{\bm E} = \left\{E_0, 0, 0 \right\}^\top
\end{eqnarray}
much like the global boundary condition itself. Then, clearly, if Eq. (\ref{E9}) is a solution, then due to linearity it is also the unique solution.

As for the interpretation of the boundary conditions, two possibilities exist.
(a) To understand the problem as two-dimensional, straight and extending between minus to plus infinity, with prescribed values there.
(b) Alternatively, to view the system as a ring with a large radius, relatively to the cross-sectional dimensions.
In this scenario, an external field would be tangential to the ring and the material field would be equal to the external field,
due to tangential component electric field continuity. 

From the engineering perspective this would actually mean that the external field
would have a large-scale gradient, such that a net current would be created in the composite fiber-reinforced ring.

The external electric field gradient would have to be large enough to create finite current in the ring, yet small enough to assume
approximately uniform field distribution within the cross-section of the ring. It is in this perspective that one may suggest an electric field gradient sensing device,
with the accompanying computational apparatus, which would be based on the derivation of the magneto-thermo-elastic response, as obtained in the following.

\section{The Magnetic Field}
\label{Sect4}

The magnetic field intensity ${\bm H}$ is governed by Eq. (\ref{E1}) in conjunction with Eq. (\ref{E4}).
Let ${\bm H}$ be decomposed into homogeneous ${\bm H}^h$ and particular ${\bm H}^p$ parts.
These parts satisfy: 
\begin{eqnarray} \label{M1}
\nabla \cdot {\bm H}^h &=& {0}, \ \ \ \ \nabla \times {\bm H}^h = {\bm 0} \nonumber \\ 
\nabla \cdot {\bm H}^p &=& {0}, \ \ \ \ \nabla \times {\bm H}^p = {\bm J} = \sigma {\bm E}  
\end{eqnarray}
It follows that ${\bm H}^h$ can be derived from a harmonic magnetic potential ${\phi}^h$ such that ${\bm H}^h = - \nabla \phi^h$.
It can be verified that ${\bm H}^p$ within subcell $(\beta \gamma)$ can be uniquely satisfied by:
\begin{eqnarray} \label{M2}
{\bm H}^{p(\beta \gamma)} = \frac{\sigma^{(\beta \gamma)} E_0}{2} \left\{ 0,-y^{(\gamma)}_3, y^{(\beta)}_2 \right\}^T
\end{eqnarray}
This is the Amp\`ere solution inside an electric current-bearing wire.
The harmonic magnetic potential can be generated as a four-terms expansion, consistent of harmonic eigenfunctions of the Laplace equation for $\phi^h$: 
\begin{eqnarray} \label{M3}
\phi^{h(\beta \gamma)} &=& \bigg[ \hat h_1 \sin  \left( \sqrt{c_1} y^{(\beta)}_2 \right) \cosh \left( \sqrt{c_1} y^{(\gamma)}_3 \right)
                               +  \hat h_2 \cos  \left( \sqrt{c_1} y^{(\beta)}_2 \right) \sinh \left( \sqrt{c_1} y^{(\gamma)}_3 \right) \nonumber \\
                       &+&        \hat h_3 \sinh \left( \sqrt{c_2} y^{(\beta)}_2 \right) \cos  \left( \sqrt{c_2} y^{(\gamma)}_3 \right)
                               +  \hat h_4 \cosh \left( \sqrt{c_2} y^{(\beta)}_2 \right) \sin  \left( \sqrt{c_2} y^{(\gamma)}_3 \right) \bigg]^{(\beta \gamma)} 
\end{eqnarray}

The reason for which the number of free coefficients in the expansion in (\ref{M3}) is four is as follows.
Since the aim of the present work is to enhance the micromechanical HOT method to include ponderomotive force effects,
and the continuity at the faces of a subcell is enforced in the average sense,
clearly, for a 2D subcell, 4 independent coefficients are needed to enforce the 4 interface conditions on the faces/edges.
Since the magnetic equations are first order in $\bm{H}$, only one condition is needed at one edge of the subcell for each of the two vector components of $\bm{H}$,
and that is for every one of the two spatial directions.
This gives a total of 4 conditions, which due to linearity correspond to 4 necessary free coefficients in a feasible expansion for the harmonic potential.
In order for the coefficients to be extractable from the interface conditions, the 4 eigenfunctions have to be orthogonal.
This can only be achieved if 2D Cartesian harmonic functions are chosen -- hence the function choice in Eq. (\ref{M3}).

The symmetry between the two spatial directions suggests that the trigonometric/hyperbolic function choice should be symmetric between the
two directions. Moreover, since two trigonometric functions are needed, symmetry and the fact that the interface conditions are integral,
leads to the fact that the entire expansion be represented by the first function, and the other three be complementary through odd/even and trigonometric/hyperbolic flips.
Hence, there are only two possibilities for the first eigenfunction, namely $\sin\sinh$ and $\sin\cosh$.
It so happens that the first choice leads to a coupled scheme, which rigorous scheme stability analysis as well as numeric examination show to be algorithmically unstable.
The second choice, on the other hand, produces a formally stable scheme, which also decouples, enabling the analytical solution of the entire magnetic problem, as shown next. 

In a discretized scheme the integral continuity conditions are the physically correct ones, since only integral continuity conditions allow the fulfillment of the
Amp\`ere and magnetic Gauss laws over arbitrary loops along subcell lines within the domain.
Consequently, the proposed scheme is the only possible first order consistent and stable discrete integration scheme for the magnetic equation with a given transverse current in a composite. 

The subcell interface continuity equations corresponding to the potential given in Eq. (\ref{M3}) are obtained by integrating
${\bm H}^{(\beta \gamma)} = - \nabla \phi^{h(\beta \gamma)} + {\bm H}^{p(\beta \gamma)}$, 
in conjunction with Eq. (\ref{M3}), over the subcell edges (the integrated values needed for interface continuity resulting in satisfaction of the integral Amp\`ere and Gauss laws for a loop
in the discretized specimen). Consequently, the coefficients $\hat{h}^{(\beta \gamma)}_1,...,\hat{h}^{(\beta \gamma)}_4$ can be established as follows:

\begin{eqnarray} \label{M4}
-2 \left[ \begin{array}{cccc}
a & 0  & b & 0   \\
0 & c  & 0 & d   \\
c & 0  & d & 0   \\
0 & a  & 0 & b   \\
\end{array} \right]^{(\beta \gamma)} \
\left[ \begin{array}{c}
\hat h_1 \\
\hat h_2 \\
\hat h_3 \\
\hat h_4 \end{array} \right]^{(\beta \gamma)}  
+ \frac{\sigma^{(\beta \gamma)} E_0}{4}  \left[ \begin{array}{c}
 - h_{\beta}  l_{\gamma} \\
   h_{\beta}  l_{\gamma} \\
   0                         \\
   0                         \\
\end{array} \right]  &=&
\left[ \begin{array}{c}
{\bar H}^{+(3)}_2 \\
{\bar H}^{+(2)}_3 \\
{\bar H}^{+(2)}_2 \\
{\bar H}^{+(3)}_3 \end{array} \right]^{(\beta \gamma)} 
\end{eqnarray}
where the superscript and subscript on $\bar H$ denote the normal to the surface and component, respectively, i.e.,
\begin{eqnarray} \label{M40}
\bar {\bm H}^{\pm (3)(\beta \gamma)} &=& \int^{h_{\beta }/2}_{-h_{\beta }/2} {\bm H}^{(\beta \gamma)} \left( y^{(\beta)}_2,  y^{(\gamma)}_3 = \pm l_{\gamma} /2 \right) dy^{(\beta )}_2 \nonumber \\
\bar {\bm H}^{\pm (2)(\beta \gamma)} &=& \int^{l_{\gamma}/2}_{-l_{\gamma}/2} {\bm H}^{(\beta \gamma)} \left( y^{(\beta)}_2 = \pm h_{\beta} /2,  y^{(\gamma)}_3  \right) dy^{(\gamma)}_3
\end{eqnarray}
and  where
\begin{eqnarray} 
a^{(\beta \gamma)} = \sin  \left( {\sqrt{c^{(\beta \gamma)}_1} h_{\beta}/2} \right) \cosh \left( {\sqrt{c^{(\beta \gamma)}_1} l_{\gamma}/2} \right),  
b^{(\beta \gamma)}  =  \sinh \left( {\sqrt{c^{(\beta \gamma)}_2} h_{\beta}/2} \right) \cos  \left( {\sqrt{c^{(\beta \gamma)}_2} l_{\gamma}/2} \right)  \nonumber \\
c^{(\beta \gamma)} = \cos  \left( {\sqrt{c^{(\beta \gamma)}_1} h_{\beta}/2} \right) \sinh \left( {\sqrt{c^{(\beta \gamma)}_1} l_{\gamma}/2} \right),
d^{(\beta \gamma)}  =  \cosh \left({\sqrt{c^{(\beta \gamma)}_2} h_{\beta}/2} \right) \sin  \left( {\sqrt{c^{(\beta \gamma)}_2} l_{\gamma}/2} \right)  
\end{eqnarray}

The solution for the coefficients is given by: 
\begin{eqnarray} \label{M5}
\left[ \begin{array}{c}
\hat h_1 \\
\hat h_2 \\
\hat h_3 \\
\hat h_4 \end{array} \right]^{(\beta \gamma)}  
= \frac{1}{2(bc-ad)^{(\beta \gamma)}}
\left[ \begin{array}{cccc}
d  & 0  & -b & 0   \\
0  & -b & 0  & d   \\
-c & 0  & a  & 0   \\
0  & a  & 0 & -c   \\
\end{array} \right]^{(\beta \gamma)} \
\left[ \begin{array}{c}
{\bar H}^{+(3)}_2 + \frac{\sigma E_0}{4} h_{\beta} l_{\gamma}  \\
{\bar H}^{+(2)}_3 - \frac{\sigma E_0}{4} h_{\beta} l_{\gamma}  \\
{\bar H}^{+(2)}_2 \\
{\bar H}^{+(3)}_3 \end{array} \right]^{(\beta \gamma)} 
\end{eqnarray}

It can be easily verified that the regularity condition requires that:
$\pi /4 < \sqrt{c^{(\beta \gamma)}_1}  h_{\beta}  / 2 < \pi/2$ and  
$\pi /4 < \sqrt{c^{(\beta \gamma)}_2}  l_{\gamma} / 2 < \pi/2$.

This can be satisfied by the choice: $c^{(\beta \gamma)}_1 = 9 \pi^2 / [16  (h^{(\beta)})^2 ], c^{(\beta \gamma)}_2 =9 \pi^2 / [16  (l^{(\gamma)})^2 ]$.   

The relations between the integrated opposite subcell edges provide:

\begin{eqnarray} \label{M6}
\left[ \begin{array}{c}
{\bar H}^{-(3)}_2  \\
{\bar H}^{-(2)}_3  \\
{\bar H}^{-(2)}_2 \\
{\bar H}^{-(3)}_3 \end{array} \right]^{(\beta \gamma)} 
= \left[ \begin{array}{c}
{\bar H}^{+(3)}_2 + \frac{\sigma E_0}{2} h_{\beta} l_{\gamma}  \\
{\bar H}^{+(2)}_3 - \frac{\sigma E_0}{2} h_{\beta} l_{\gamma}  \\
{\bar H}^{+(2)}_2          \\
{\bar H}^{+(3)}_3    \end{array} \right]^{(\beta \gamma)} 
\end{eqnarray}

Face-integrated continuity of interfaces, along with Eq. (\ref{M6}) yields:

\begin{eqnarray} \label{M7}
\left[ \begin{array}{c}
{\bar H}^{+(3)(\beta, \gamma+1)}_2  \\
{\bar H}^{+(2)(\beta+1, \gamma)}_3  \\
\mu^{(\beta+1, \gamma)}_r {\bar H}^{+(2)(\beta+1, \gamma)}_2 \\
\mu^{(\beta, \gamma+1)}_r {\bar H}^{+(3)(\beta, \gamma+1)}_3 \end{array} \right]
= \left[ \begin{array}{c}
{\bar H}^{+(3)(\beta \gamma)}_2  \\
{\bar H}^{+(2)(\beta \gamma)}_3  \\
\mu^{(\beta \gamma)}_r {\bar H}^{+(2)(\beta \gamma)}_2 \\
\mu^{(\beta \gamma)}_r {\bar H}^{+(3)(\beta \gamma)}_3 \end{array} \right]
+ \left[ \begin{array}{c}
- \frac{ E_0}{2} h_{\beta}   l_{\gamma+1} \sigma^{(\beta, \gamma+1)}  \\
  \frac{ E_0}{2} h_{\beta+1} l_{\gamma}   \sigma^{(\beta+1, \gamma)}  \\
  0          \\
  0    \end{array} \right]
\end{eqnarray}

Solving the recursion equation (\ref{M7}) analytically gives:

\begin{eqnarray} \label{M8}
\left[ \begin{array}{c}
{\bar H}^{+(3)(\beta \gamma)}_2  \\
{\bar H}^{+(2)(\beta \gamma)}_3  \\
{\bar H}^{+(2)(\beta \gamma)}_2 \\
{\bar H}^{+(3)(\beta \gamma)}_3 \end{array} \right]
= \left[ \begin{array}{c}
{\hat H}^{(\beta)}_{23}  - \frac{E_0}{2} \sum^{\gamma}_{\gamma'=1} h_{\beta}  l_{\gamma'} \sigma^{(\beta \gamma')}   \\
{\hat H}^{(\gamma)}_{32} + \frac{E_0}{2} \sum^{\beta}_{\beta'=1}   h_{\beta'} l_{\gamma}  \sigma^{(\beta' \gamma)}  \\
\frac{1}{\mu^{(\beta \gamma)}_r}  {\tilde H}^{-(2)(1, \gamma)}_2    \\
\frac{1}{\mu^{(\beta \gamma)}_r}  {\tilde H}^{-(3)(\beta, 1)}_3   \end{array} \right]
= \left[ \begin{array}{c}
{\hat H}^{(\beta)}_{23}  - \frac{E_0}{2} \sum^{\gamma}_{\gamma'=1} h_{\beta}  l_{\gamma'} \sigma^{(\beta \gamma')}   \\
{\hat H}^{(\gamma)}_{32} + \frac{E_0}{2} \sum^{\beta}_{\beta'=1}   h_{\beta'} l_{\gamma}  \sigma^{(\beta' \gamma)}  \\
\frac{1}{\mu^{(\beta \gamma)}_r}  {\tilde H}^{+(2)(N_{\beta}, \gamma)}_2    \\
\frac{1}{\mu^{(\beta \gamma)}_r}  {\tilde H}^{+(3)(\beta, N_{\gamma})}_3   \end{array} \right]
\end{eqnarray}

The last relation in (\ref{M8}) implies that components 3 and 4 can be also taken at the opposite global boundary,
since the magnetic flux densities should be uniform in the direction in which they
point. Thus symmetric global boundary conditions are assumed for the normal magnetic flux density components, i.e., a symmetric specimen and hence skew-symmetric tangential
components' global boundary conditions. 

The assumption is reasonable, since normal magnetic flux components penetrate
the composite remaining unaltered passing through different constituents, due to the fact that normal magnetic flux
continuity across interfaces is permeability-independent (unlike for tangential magnetic flux components).

Next, in order to obtain the unknown single-index functions in the first two components in the right-hand-side of Eq. (\ref{M8}), we write them explicitly
for the two opposite global boundaries (for each of the first two components there), and then solve for the unknowns, using skew-symmetry of
the tangential field for a symmetric-cross-section conducting wire. 

This yields: 
\begin{eqnarray} \label{M9}
{\hat H}^{(\gamma)}_{32} = - \frac{E_0}{4} l_{\gamma} \sum^{N_{\beta}}_{\beta = 1}   h_{\beta}  \sigma^{(\beta \gamma)} \ \ \ , \ \ \ 
{\hat H}^{(\beta)}_{23}  =   \frac{E_0}{4} h_{\beta } \sum^{N_{\gamma}}_{\gamma = 1} l_{\gamma} \sigma^{(\beta \gamma)}
\end{eqnarray}

Substituting these relations in Eq. (\ref{M8}), we get the face-integrated field components explicitly, depending only on the global boundary conditions: 
\begin{eqnarray} \label{M10}
\left[ \begin{array}{c}
{\bar H}^{+(3)}_2  \\
{\bar H}^{+(2)}_3  \\
{\bar H}^{+(2)}_2 \\
{\bar H}^{+(3)}_3 \end{array} \right]^{(\beta \gamma)} 
=\left[ \begin{array}{c}
 \frac{E_0}{4} \left(\sum^{N_{\gamma}}_{\gamma=1}   h_{\beta}  l_{\gamma} \sigma^{(\beta \gamma)} - 2 \sum^{\gamma}_{\gamma'=1} h_{\beta} l_{\gamma'} \sigma^{(\beta \gamma')}\right)  \\
 \frac{E_0}{4} \left(2 \sum^{\beta}_{\beta'=1} h_{\beta'} l_{\gamma} \sigma^{(\beta' \gamma)} - \sum^{N_{\beta}}_{\beta=1} h_{\beta} l_{\gamma}  \sigma^{(\beta \gamma)}  \right) \\
          \frac{1}{\mu^{(\beta \gamma)}_r}{\tilde H}^{-(2)(1, \gamma)}_2        \\
          \frac{1}{\mu^{(\beta \gamma)}_r}{\tilde H}^{-(3)(\beta ,1)}_3         \end{array} \right]
\end{eqnarray}

Hence, the final expressions of the coefficients $\hat h^{(\beta \gamma)}_1,..., \hat h^{(\beta \gamma)}_4$ take the form: 
\begin{eqnarray} \label{M11}
\hat h^{(\beta \gamma)}_1 &=& \frac{ \frac{E_0}{4}h_{\beta} d^{(\beta \gamma)} \left[ \sum^{N_{\gamma}}_{\gamma=1} l_{\gamma} \sigma^{(\beta \gamma)}
                       - 2 \sum^{\gamma}_{\gamma'=1} l_{\gamma'} \sigma^{(\beta \gamma')}
                       + l_{\gamma} \sigma^{(\beta \gamma)} \right] - b^{(\beta \gamma)} {\tilde H}^{-(2)(1, \gamma)}_2 / \mu^{(\beta \gamma)}_r}
                       {2(bc-ad)^{(\beta \gamma)}} \nonumber \\  
\hat h^{(\beta \gamma)}_2 &=& \frac{ \frac{E_0}{4}l_{\gamma} b^{(\beta \gamma)} \left[ \sum^{N_{\beta}}_{\beta=1} h_{\beta} \sigma^{(\beta \gamma)}
                       - 2 \sum^{\beta}_{\beta'=1} h_{\beta'} \sigma^{(\beta' \gamma)}
                       + h_{\beta} \sigma^{(\beta \gamma)} \right] + d^{(\beta \gamma)} {\tilde H}^{-(3)(\beta, 1)}_3 / \mu^{(\beta \gamma)}_r}
                       {2(bc-ad)^{(\beta \gamma)}} \nonumber \\  
\hat h^{(\beta \gamma)}_3 &=&  \frac{- \frac{E_0}{4}h_{\beta} c^{(\beta \gamma)} \left[ \sum^{N_{\gamma}}_{\gamma=1} l_{\gamma} \sigma^{(\beta \gamma)}
                       - 2 \sum^{\gamma}_{\gamma'=1} l_{\gamma'} \sigma^{(\beta \gamma')}
                       + l_{\gamma} \sigma^{(\beta \gamma)} \right] + a^{(\beta \gamma)} {\tilde H}^{-(2)(1, \gamma)}_2 / \mu^{(\beta \gamma)}_r}
                       {2(bc-ad)^{(\beta \gamma)}} \nonumber \\  
\hat h^{(\beta \gamma)}_4 &=&\frac{- \frac{E_0}{4}l_{\gamma} a^{(\beta \gamma)} \left[ \sum^{N_{\beta}}_{\beta=1} h_{\beta} \sigma^{(\beta \gamma)}
                       - 2 \sum^{\beta}_{\beta'=1} h_{\beta'} \sigma^{(\beta' \gamma)}
                       + h_{\beta} \sigma^{(\beta \gamma)} \right] - c^{(\beta \gamma)} {\tilde H}^{-(3)(\beta, 1)}_3 / \mu^{(\beta \gamma)}_r}
                       {2(bc-ad)^{(\beta \gamma)}} \nonumber \\  
\end{eqnarray}
where the normal components of the global boundary conditions for magnetic field intensity are obtained by integrating the Biot-Savart law both over each conducting subcell and each global
boundary subcell edge: 
\begin{eqnarray} \label{M12}
{\tilde H}^{-(2)(1, \gamma)}_2 &=& \frac{E_0}{4 \pi l_{\gamma}} \sum^{N_{\gamma}}_{\gamma'=1} \sum^{N_{\beta}}_{\beta'=1} \sigma^{(\beta' \gamma')}
                    \bigg[ 2 I_2 \left( \sum^{\beta'  }_{n=1} h_{n}, \bar \xi_3, \bar \xi_3 + l_{\gamma} \right)
                          -2 I_2 \left( \sum^{\beta'-1}_{n=1} h_{n}, \bar \xi_3, \bar \xi_3 + l_{\gamma} \right) \nonumber \\
                       &+& 2 I_2 \left( \sum^{\beta'-1}_{n=1} h_{n}, \bar \xi_3-l_{\gamma}, \bar \xi_3  \right)
                        -  2 I_2 \left( \sum^{\beta'  }_{n=1} h_{n}, \bar \xi_3-l_{\gamma}, \bar \xi_3  \right)  \nonumber \\
   &+& \left(\sum^{\beta'  }_{n=1} h_{n} \right) I_1 \left( \sum^{\beta'  }_{n=1} h_{n}, \bar \xi_3, \bar \xi_3+l_{\gamma}  \right)
    +  \left(\sum^{\beta'-1}_{n=1} h_{n} \right) I_1 \left( \sum^{\beta'-1}_{n=1} h_{n}, \bar \xi_3-l_{\gamma}, \bar \xi_3  \right) \nonumber \\
   &-& \left(\sum^{\beta'  }_{n=1} h_{n} \right) I_1 \left( \sum^{\beta'  }_{n=1} h_{n}, \bar \xi_3-l_{\gamma}, \bar \xi_3  \right)
    -  \left(\sum^{\beta'-1}_{n=1} h_{n} \right) I_1 \left( \sum^{\beta'-1}_{n=1} h_{n}, \bar \xi_3, \bar \xi_3+l_{\gamma}  \right) \bigg]
\end{eqnarray}
\begin{eqnarray} \label{M13}
{\tilde H}^{-(3)(\beta, 1)}_3 &=& \frac{E_0}{4 \pi h_{\beta}} \sum^{N_{\gamma}}_{\gamma'=1} \sum^{N_{\beta}}_{\beta'=1} \sigma^{(\beta' \gamma')}
                    \bigg[ 2 I_2 \left( \sum^{\gamma'   }_{n=1}   l_{n}, \bar \xi_2, \bar \xi_2 + h_{\beta} \right)
                          -2 I_2 \left( \sum^{\gamma'-1 }_{n=1}   l_{n}, \bar \xi_2, \bar \xi_2 + h_{\beta} \right) \nonumber \\
                       &+& 2 I_2 \left( \sum^{\gamma'-1 }_{n=1}   l_{n}, \bar \xi_2-h_{\beta }, \bar \xi_2  \right)
                        -  2 I_2 \left( \sum^{\gamma'   }_{n=1}   l_{n}, \bar \xi_2-h_{\beta }, \bar \xi_2  \right)  \nonumber \\
   &+& \left(\sum^{\gamma'  }_{n=1} l_{n} \right) I_1 \left( \sum^{\gamma'  }_{n=1} l_{n}, \bar \xi_2, \bar \xi_2+h_{\beta}  \right)
    +  \left(\sum^{\gamma'-1}_{n=1} l_{n} \right) I_1 \left( \sum^{\gamma'-1}_{n=1} l_{n}, \bar \xi_2-h_{\beta }, \bar \xi_2  \right) \nonumber \\
   &-& \left(\sum^{\gamma'  }_{n=1} l_{n} \right) I_1 \left( \sum^{\gamma'  }_{n=1} l_{n}, \bar \xi_2-h_{\beta }, \bar \xi_2  \right)
    -  \left(\sum^{\gamma'-1}_{n=1} l_{n} \right) I_1 \left( \sum^{\gamma'-1}_{n=1} l_{n}, \bar \xi_2, \bar \xi_2+h_{\beta}  \right) \bigg]
\end{eqnarray}
\begin{eqnarray} \label{M14}
\bar \xi_2 = \sum^{\beta'-1}_{n=1} h_{n} -  \sum^{\beta-1}_{n=1} h_{n}, \ \ \ \ \bar \xi_3 = \sum^{\gamma'-1}_{n=1} l_{n} - \sum^{\gamma-1}_{n=1} l_{n}
\end{eqnarray}
\begin{eqnarray} \label{M15}
I_1 (a,x_1,x_2) &=& \int^{x_2}_{x_1} \ln (a^2+x^2) dx = -2 (x_2-x_1) + 2 a \left[\tan^{-1}\left(\frac{x_2}{a}\right) - \tan^{-1}\left(\frac{x_1}{a}\right) \right] \nonumber \\
                &+& \ln \left[ (a^2 + x^2_2)^{x_2} \right] -  \ln \left[ (a^2 + x^2_1)^{x_1} \right]    \nonumber \\
I_2 (a,x_1,x_2) &=& \int^{x_2}_{x_1} x \tan^{-1}\left(\frac{a}{x}\right) dx = \frac{a}{2} (x_2-x_1) - \frac{a^2}{2} \left[\tan^{-1}\left(\frac{x_2}{a}\right)-\tan^{-1}\left(\frac{x_1}{a}\right) \right] \nonumber \\
                &+& \frac{1}{2} \left[ x^2_2 \tan^{-1} \left(\frac{a}{x_2}\right) - x^2_1 \tan^{-1} \left(\frac{a}{x_1}\right) \right]
\end{eqnarray}

Just like in \cite{Librescu2003}, the global boundary conditions are presently obtained by summing up the contributions of all the current-bearing fibers,
along with the Biot-Savart law. The difference is that here, two additional integrations were performed,
one over each conducting fiber area, to account for the integral contribution without approximation,
and the other is integration along a global boundary subcell edge, which is needed as the boundary conditions are enforced in the integral sense, for reasons explained earlier.

Finally, the magnetic field intensity distribution components in subcell $(\beta \gamma)$ are given by:
\begin{eqnarray} \label{M16}
H^{(\beta \gamma)}_1 (y^{(\beta)}_2, y^{(\gamma)}_3 ) &=& 0  \nonumber \\ 
H^{(\beta \gamma)}_2 (y^{(\beta)}_2, y^{(\gamma)}_3 ) &=& \bigg[
  - \sqrt{c_1} \hat h_1 \cos  \left(\sqrt{c_1} y^{(\beta)}_2 \right) \cosh \left(\sqrt{c_1} y^{(\gamma)}_3 \right)
  + \sqrt{c_1} \hat h_2 \sin  \left(\sqrt{c_1} y^{(\beta)}_2 \right) \sinh \left(\sqrt{c_1} y^{(\gamma)}_3 \right)  \nonumber \\   
&-& \sqrt{c_2} \hat h_3 \cosh \left(\sqrt{c_2} y^{(\beta)}_2 \right) \cos  \left(\sqrt{c_2} y^{(\gamma)}_3 \right)              
  - \sqrt{c_2} \hat h_4 \sinh \left(\sqrt{c_2} y^{(\beta)}_2 \right) \sin  \left(\sqrt{c_2} y^{(\gamma)}_3 \right) \nonumber \\ 
&-& \frac{1}{2}E_0 \sigma y^{(\gamma)}_3 \bigg]^{(\beta \gamma)} 
\end{eqnarray}

\begin{eqnarray} \label{M16b}          
H^{(\beta \gamma)}_3 (y^{(\beta)}_2, y^{(\gamma)}_3 ) &=& \bigg[
  - \sqrt{c_1} \hat h_1 \sin  \left(\sqrt{c_1} y^{(\beta)}_2 \right) \sinh \left(\sqrt{c_1} y^{(\gamma)}_3 \right)
  - \sqrt{c_1} \hat h_2 \cos  \left(\sqrt{c_1} y^{(\beta)}_2 \right) \cosh \left(\sqrt{c_1} y^{(\gamma)}_3 \right)  \nonumber \\   
&+& \sqrt{c_2} \hat h_3 \sinh \left(\sqrt{c_2} y^{(\beta)}_2 \right) \sin  \left(\sqrt{c_2} y^{(\gamma)}_3 \right)              
  - \sqrt{c_2} \hat h_4 \cosh \left(\sqrt{c_2} y^{(\beta)}_2 \right) \cos  \left(\sqrt{c_2} y^{(\gamma)}_3 \right)  \nonumber \\             
&+& \frac{1}{2}E_0 \sigma y^{(\beta)}_2 \bigg]^{(\beta \gamma)}                
\end{eqnarray}

With the established electric and magnetic field intensities, $\bm{E}$ and ${\bm H}^{(\beta \gamma)}$, in each subcell,
the ponderomotive force variation in the subcell, as given in Eq. (\ref{E8}), can be expressed as follows: 
\begin{eqnarray} \label{M17}
{\bm F}^{(\beta \gamma)}(y^{(\beta)}_2, y^{(\gamma)}_3 )= \mu_0 \left[ \mu_r {\bm H}^\top \nabla {\bm H} - \left( \nabla {\bm H}^\top \right) {\bm H} \right]^{(\beta \gamma)}
\end{eqnarray}

\section{The Thermal Field}
\label{Sect5}

In the following, an approximate solution for the temperature distribution in the composite is established using a hybrid HOT -- finite difference approach.
From Eqs. (\ref{E2}), (\ref{E5}a), (\ref{E6}) and (\ref{E9}), the temperature field within each subcell should satisfy Poisson's equation:
\begin{eqnarray} \label{T1}
\nabla^2 \theta = - \frac{\sigma}{\kappa} E^2_0
\end{eqnarray}

Let the temperature $\theta^{(\beta \gamma)}$ within subcell $(\beta \gamma)$ be decomposed into the homogeneous and particular parts:
\begin{eqnarray} \label{T0}
\theta^{(\beta \gamma)}=\theta^{h(\beta \gamma)} + \theta^{p(\beta \gamma)}
\end{eqnarray}
such that
\begin{eqnarray} \label{T2}
\nabla^2 \theta^{h(\beta \gamma)} = 0, \ \ \ \  \nabla^2 \theta^{p(\beta \gamma)} = - \frac{\sigma^{(\beta \gamma)}}{\kappa^{(\beta \gamma)}} E^2_0
\end{eqnarray}
A solution to Eq. (\ref{T2}b) is:
\begin{eqnarray} \label{T3}
\theta^{p(\beta \gamma)} = -\frac{E^2_0}{4} \frac{\sigma^{(\beta \gamma)}}{\kappa^{(\beta \gamma)}} \left[(y^{(\beta)}_2)^2 + (y^{(\gamma)}_3)^2 \right]
\end{eqnarray}
The homogeneous solution can be chosen in the form:
\begin{eqnarray} \label{T4}
\theta^{h(\beta \gamma)} &=& \bigg[ \hat \theta_1 \cos  \left(\sqrt{\lambda_1} y^{(\beta)}_2 \right) \cosh \left(\sqrt{\lambda_1} y^{(\gamma)}_3 \right)
                          +         \hat \theta_2 \cos  \left(\sqrt{\lambda_1} y^{(\beta)}_2 \right) \sinh \left(\sqrt{\lambda_1} y^{(\gamma)}_3 \right) \nonumber \\
                         &+&        \hat \theta_3 \sinh \left(\sqrt{\lambda_2} y^{(\beta)}_2 \right) \cos  \left(\sqrt{\lambda_2} y^{(\gamma)}_3 \right)
                          +         \hat \theta_4 \cosh \left(\sqrt{\lambda_2} y^{(\beta)}_2 \right) \cos  \left(\sqrt{\lambda_2} y^{(\gamma)}_3 \right) \bigg]^{(\beta \gamma)} 
\end{eqnarray}

This expression is constructed in a manner similar to the one used in the magnetic case -- four linearly independent harmonic eigenfunctions,
corresponding to to the four local boundary conditions for each subcell, two on each edge, as expected for a Poisson equation.
The only difference is that this time there are two 'even-even' functions, in addition to two 'odd-odd' functions.
This is due to the fact that in this case one condition is for heat flux continuity, which is an integral condition,
but the other condition is for temperature continuity, which has no integral version and is a macroscopically local requirement. Moreover, no differentiation is required, just value setting. 

These two features lead to the choice of two 'even-even' eigenfunctions, as a means to correctly represent bidirectional thermal self-coupling in the resulting integration scheme.
Now, evaluating the temperature at the centers of the edges $y^{(\beta)}_2 = h_{\beta} / 2, y^{(\gamma)}_3 =0$ and $y^{(\beta)}_2 =0, y^{(\gamma)}_3 = l_{\gamma} / 2$,
and, in addition, integrating the normal heat flux densities along these edges, yields, in conjunction with Eq. (\ref{T0}), the following relation: 
\begin{eqnarray} 
\label{T5}
 \left \lbrace\left[ \begin{array}{cccc}
 2A    & 0      & -2B    & -2D      \\
-2A    & -2C    & 0      &  2D      \\
\bar A & 0      & \bar B & \bar F   \\
\bar G & \bar C & 0      & \bar D   \\
\end{array} \right]
\left[ \begin{array}{c}
\hat \theta_1 \\
\hat \theta_2 \\
\hat \theta_3 \\
\hat \theta_4 \end{array} \right]\right\rbrace^{(\beta \gamma)}
+ \frac{\sigma^{(\beta \gamma)} E^2_0}{4 \kappa^{(\beta \gamma)}}  \left[ \begin{array}{c}
   h_{\beta} l_{\gamma}       \\
   h_{\beta} l_{\gamma}       \\
  - \frac{1}{4} h^2_{\beta}    \\
  - \frac{1}{4} l^2_{\gamma}   \\
\end{array} \right] 
 =
\left[ \begin{array}{c}
\frac{1}{\kappa^{(\beta \gamma)}} {\bar q}^{+(\beta \gamma)}_2 \\
\frac{1}{\kappa^{(\beta \gamma)}} {\bar q}^{+(\beta \gamma)}_3 \\
{\theta}|_{ y^{(\beta)}_2 = \frac{h_{\beta}}{2}, y^{(\gamma)}_3 = 0}      \\
{\theta}|_{y^{(\beta)}_2 = 0, y^{(\gamma)}_3 = \frac{l_{\gamma}}{2}}  \end{array}  \right]
\end{eqnarray}
with these auxiliary definitions employed:
\begin{eqnarray} \label{T6}
A^{(\beta \gamma)} \triangleq \sin  \left( \sqrt{\lambda^{(\beta \gamma)}_1} \frac{h_{\beta}}{2} \right) \sinh \left( \sqrt{\lambda^{(\beta \gamma)}_1} \frac{l_{\gamma}}{2} \right), \ \ \ \ 
B^{(\beta \gamma)} \triangleq \cosh \left( \sqrt{\lambda^{(\beta \gamma)}_2} \frac{h_{\beta}}{2} \right) \sin  \left( \sqrt{\lambda^{(\beta \gamma)}_2} \frac{l_{\gamma}}{2} \right) \nonumber \\
C^{(\beta \gamma)} \triangleq \sin  \left( \sqrt{\lambda^{(\beta \gamma)}_1} \frac{h_{\beta}}{2} \right) \cosh \left( \sqrt{\lambda^{(\beta \gamma)}_1} \frac{l_{\gamma}}{2} \right), \ \ \ \ 
D^{(\beta \gamma)} \triangleq \sinh \left( \sqrt{\lambda^{(\beta \gamma)}_2} \frac{h_{\beta}}{2} \right) \sin  \left( \sqrt{\lambda^{(\beta \gamma)}_2} \frac{l_{\gamma}}{2} \right) \nonumber \\
\bar A^{(\beta \gamma)} \triangleq \cos  \left( \sqrt{\lambda^{(\beta \gamma)}_1} \frac{h_{\beta}}{2} \right), \ \ \ \
\bar B^{(\beta \gamma)} \triangleq \sinh \left( \sqrt{\lambda^{(\beta \gamma)}_2} \frac{h_{\beta}}{2} \right), \ \ \ \ 
\bar C^{(\beta \gamma)} \triangleq \sinh \left( \sqrt{\lambda^{(\beta \gamma)}_1} \frac{l_{\gamma}}{2} \right)                        \nonumber \\
\bar D^{(\beta \gamma)} \triangleq \cos  \left( \sqrt{\lambda^{(\beta \gamma)}_2} \frac{l_{\gamma}}{2} \right), \ \ \ \
\bar F^{(\beta \gamma)} \triangleq \cosh \left( \sqrt{\lambda^{(\beta \gamma)}_2} \frac{h_{\beta}}{2} \right), \ \ \ \ 
\bar G^{(\beta \gamma)} \triangleq \cosh \left( \sqrt{\lambda^{(\beta \gamma)}_1} \frac{l_{\gamma}}{2} \right) 
\end{eqnarray}

A regularity condition, corresponding to a non-vanishing determinant of the matrix in Eq. (\ref{T5}), is satisfied by the symmetric choice:
$\sqrt{\lambda^{(\beta \gamma)}_1} h_{\beta}/2 = \sqrt{\lambda^{(\beta \gamma)}_2} l_{\gamma}/2 = \pi/4$, for which the determinant, $\Delta$, of the matrix in Eq. (\ref{T6}), 
satisfies the inequality: $\Delta \le -3.7205$, guaranteeing regularity. 

Next, evaluating the temperature at the centers of the opposite pair of adjacent edges, $y^{(\beta)}_2 = - h_{\beta} / 2, y^{(\gamma)}_3 =0$ and $y^{(\beta)}_2 =0, y^{(\gamma)}_3 = - l_{\gamma} / 2$,
and, in addition, integrating  the normal heat flux densities along these edges, yields, in conjunction with Eq. (\ref{T4}), the following, second set of relations:
\begin{eqnarray} \label{T7}
\left\lbrace \left[ \begin{array}{cccc}
-2A    & 0        & -2B    &  2D      \\
 2A    & -2C      & 0      & -2D      \\
\bar A & 0        &-\bar B & \bar F   \\
\bar G & - \bar C & 0      & \bar D   \\
\end{array} \right]
\left[ \begin{array}{c}
\hat \theta_1 \\
\hat \theta_2 \\
\hat \theta_3 \\
\hat \theta_4 \end{array} \right]\right\rbrace^{(\beta \gamma)}
- \frac{\sigma^{(\beta \gamma)}E^2_0}{4 \kappa^{(\beta \gamma)}}  \left[ \begin{array}{c}
  h_{\beta} l_{\gamma}        \\
  h_{\beta} l_{\gamma}        \\
   \frac{1}{4} h^2_{\beta}    \\
   \frac{1}{4} l^2_{\gamma}   \\
\end{array} \right]
 =
\left[ \begin{array}{c}
\frac{1}{\kappa^{(\beta \gamma)}} {\bar q}^{-(\beta \gamma)}_2 \\
\frac{1}{\kappa^{(\beta \gamma)}} {\bar q}^{-(\beta \gamma)}_3 \\
{\theta}|_{y^{(\beta)}_2 =-\frac{h_{\beta}}{2}, y^{(\gamma)}_3 = 0}      \\
{\theta}|_{y^{(\beta)}_2 = 0, y^{(\gamma)}_3 =-\frac{l_{\gamma}}{2}}   \end{array}  \right]
\end{eqnarray}

Consequently, the temperature and normal heat flux 'jumps' between the opposite edges of subcell $(\beta \gamma)$, can be expressed as follows: 
\begin{eqnarray} \label{T8} 
\left[ \begin{array}{c}
\frac{1}{\kappa^{(\beta \gamma)}} \left( {\bar q}^{+(\beta \gamma)}_2 - \frac{E^2_0}{4} \sigma^{(\beta \gamma)} h_{\beta} l_{\gamma} \right)   \\
\frac{1}{\kappa^{(\beta \gamma)}} \left( {\bar q}^{+(\beta \gamma)}_3 - \frac{E^2_0}{4} \sigma^{(\beta \gamma)} h_{\beta} l_{\gamma} \right)   \\
{\theta}|_{y^{(\beta)}_2 = \frac{h_{\beta}}{2}, y^{(\gamma)}_3 = 0} + \frac{\sigma^{(\beta \gamma)}}{16 \kappa^{(\beta \gamma)}} E^2_0 h^2_{\beta}       \\
{\theta}|_{y^{(\beta)}_2 = 0, y^{(\gamma)}_3 = \frac{l_{\gamma}}{2}}+ \frac{\sigma^{(\beta \gamma)}}{16 \kappa^{(\beta \gamma)}} E^2_0 l^2_{\gamma}    \end{array}  \right]
= {\bm M}^{(\beta \gamma)}
\left[ \begin{array}{c}
\frac{1}{\kappa^{(\beta \gamma)}} \left( {\bar q}^{-(\beta \gamma)}_2 + \frac{E^2_0}{4} \sigma^{(\beta \gamma)} h_{\beta} l_{\gamma} \right)   \\
\frac{1}{\kappa^{(\beta \gamma)}} \left( {\bar q}^{-(\beta \gamma)}_3 + \frac{E^2_0}{4} \sigma^{(\beta \gamma)} h_{\beta} l_{\gamma} \right)   \\
{\theta}|_{y^{(\beta)}_2 =-\frac{h_{\beta}}{2}, y^{(\gamma)}_3 = 0} + \frac{\sigma^{(\beta \gamma)}}{16 \kappa^{(\beta \gamma)}} E^2_0 h^2_{\beta}       \\
{\theta}|_{y^{(\beta)}_2 = 0, y^{(\gamma)}_3 =-\frac{l_{\gamma}}{2}}+ \frac{\sigma^{(\beta \gamma)}}{16 \kappa^{(\beta \gamma)}} E^2_0 l^2_{\gamma}    \end{array}  \right] 
\end{eqnarray} 
with the following auxiliary matrix used:
\begin{eqnarray} \label{T9}
{\bm M}^{(\beta \gamma)} \triangleq
 \left\{ \left[ \begin{array}{cccc}
 2A    & 0        & -2B    & -2D      \\
 -2A   & -2C      & 0      &  2D      \\
\bar A & 0        & \bar B & \bar F   \\
\bar G &   \bar C & 0      & \bar D   \\
\end{array} \right] \
 \left[ \begin{array}{cccc}
-2A    & 0        & -2B    &  2D      \\
 2A    & -2C      & 0      & -2D      \\
\bar A & 0        &-\bar B & \bar F   \\
\bar G &  -\bar C & 0      & \bar D   \\
\end{array} \right]^{-1} \right\}^{(\beta \gamma)} \
\end{eqnarray}
where it can be shown that: det$\left( {\bm M}^{(\beta \gamma)} \right)$ = 1.

Rearranging Eq. (\ref{T8}), the final form of a set of $4N_{\beta}N_{\gamma}$ subcell equations relating opposite edges' values for a given subcell is obtained:
\begin{eqnarray} \label{T10}
 \left[ \begin{array}{c}
\frac{1}{\kappa^{(\beta \gamma)}}  {\bar q}^{+(\beta \gamma)}_2    \\
\frac{1}{\kappa^{(\beta \gamma)}}  {\bar q}^{+(\beta \gamma)}_3    \\
{\theta}|_{y^{(\beta)}_2 = \frac{h_{\beta}}{2}, y^{(\gamma)}_3 = 0}     \\
{\theta}|_{y^{(\beta)}_2 = 0, y^{(\gamma)}_3 = \frac{l_{\gamma}}{2}}  \end{array}  \right]
 =  {\bm M}^{(\beta \gamma)} 
\left[ \begin{array}{c}
\frac{1}{\kappa^{(\beta \gamma)}}  {\bar q}^{-(\beta \gamma)}_2    \\
\frac{1}{\kappa^{(\beta \gamma)}}  {\bar q}^{-(\beta \gamma)}_3    \\
{\theta}|_{y^{(\beta)}_2 =-\frac{h_{\beta}}{2}, y^{(\gamma)}_3 = 0}     \\
{\theta}|_{y^{(\beta)}_2 = 0, y^{(\gamma)}_3 =-\frac{l_{\gamma}}{2}}  \end{array}  \right] 
+ \frac{E^2_0}{4} \frac{\sigma^{(\beta \gamma)}}{\kappa^{(\beta \gamma)}} h_{\beta} l_{\gamma} \left\lbrace{\bm M}^{(\beta \gamma)} 
\left[ \begin{array}{c}
1    \\
1    \\
\frac{h_{\beta}}{l_{\gamma}}     \\
\frac{l_{\gamma}}{h_{\beta}}     \end{array}  \right] 
 -  
\left[ \begin{array}{c}
-1    \\
-1    \\
\frac{h_{\beta}}{l_{\gamma}}     \\
\frac{l_{\gamma}}{h_{\beta}}     \end{array}  \right] \right\rbrace
\end{eqnarray}

In addition to the above 'subcell' relations, interfacial conditions between neighboring subcells must be satisfied, as follows:
\begin{eqnarray} \label{T11}
{\bar q}^{+(\beta \gamma)}_2  &=& {\bar q}^{-(\beta+1, \gamma)}_2 \nonumber \\
{\bar q}^{+(\beta \gamma)}_3  &=& {\bar q}^{-(\beta, \gamma+1)}_3 \nonumber \\
{\theta}^{(\beta \gamma)}\left|_{y^{(\beta)}_2 =\frac{h_{\beta}}{2}, y^{(\gamma)}_3 = 0}\right. &=& {\theta}^{(\beta+1, \gamma)}\left|_{y^{(\beta+1)}_2 =-\frac{h_{\beta+1}}{2}, y^{(\gamma)}_3 = 0} \right.\nonumber \\
{\theta}^{(\beta \gamma)}\left|_{y^{(\beta)}_2 = 0, y^{(\gamma)}_3 =\frac{l_{\gamma}}{2}}\right. &=& {\theta}^{(\beta, \gamma+1)}\left|_{y^{(\beta)}_2 = 0, y^{(\gamma+1)}_3 =-\frac{l_{\gamma+1}}{2}}\right.
\end{eqnarray}

Finally, the boundary conditions at the external boundaries must be imposed. We follow here \cite{Librescu2003} by assuming convection boundary conditions, which take the form:  
\begin{eqnarray} \label{T12}
{\bar q}^{-(1, \gamma)}_2         &=& - l_{\gamma} \Lambda^{(\gamma)}_1           \ {\theta}^{(1, \gamma)}\left|_{y^{(1)}_2 =-\frac{h_{1}}{2}, y^{(\gamma)}_3 = 0} \right.      \nonumber \\
{\bar q}^{+(N_{\beta}, \gamma)}_2 &=&   l_{\gamma} \Lambda^{(\gamma)}_{N_{\beta}} \ {\theta}^{(N_{\beta},\gamma)} \left|_{y^{(N_{\beta})}_2 =\frac{h_{N_{\beta}}}{2}, y^{(\gamma)}_3 = 0} \right.  \nonumber \\
{\bar q}^{-(\beta, 1)}_3          &=& - h_{\beta}  \Lambda^{(\beta )}_1           \ {\theta}^{(\beta, 1)} \left|_{y^{(\beta)}_2 =0, y^{(1)}_3 = - \frac{l_{1}}{2}} \right.      \nonumber \\
{\bar q}^{+(\beta, N_{\gamma})}_3 &=&   h_{\beta}  \Lambda^{(\beta)}_{N_{\gamma}} \ {\theta}^{(\beta, N_{\gamma})}\left|_{y^{(\beta)}_2 =0, y^{(N_{\gamma})}_3 =  \frac{l_{N_{\gamma}}}{2}} \right.    
\end{eqnarray}
where $\Lambda^{(\gamma)}_{\beta}$ and $\Lambda^{(\beta)}_{\gamma}$ are the convection coefficients prescribed at the boundaries $X_2=\pm H/2$ and $X_3=\pm L/2$, respectively.
In general, the convection coefficient is defined by $\Lambda =  \kappa_0\text{Nu} / L_{max}$ where $\text{Nu}$, $\kappa_0$ and $L_{max}$ are
the Nusselt number, thermal conductivity of ambient air ($\kappa_0=0.024 Wm^{-1}K^{-1}$) and the length of the largest side of the specimen, respectively.

In every subcell there are 8 unknowns given by the 4 temperatures at the centers of the edges and the 4 integrals along the subcell edges of the normal heat flux densities.
Thus, there are $8 N_{\beta} N_{\gamma}$ unknown variables in the considered composite. In the same time, Eq. (\ref{T10}) provides $4 N_{\beta} N_{\gamma}$ relations.
In addition, the interfacial conditions (\ref{T11}) form $2 (N_{\beta}-1) N_{\gamma} + 2 N_{\beta} (N_{\gamma}-1)$ relations.    
Finally, the external boundary conditions, (\ref{T12}), provide another $2 (N_{\beta} + N_{\gamma}) $ relations. Consequently, there are $8 N_{\beta} N_{\gamma}$ relations, which
can be solved for the mid-edges' temperatures and total normal heat fluxes. These enable the determination of the coefficients $\hat \theta_1, ..., \hat \theta_4$, by employing
Eq. (\ref{T7}) or (\ref{T9}).
Last, the temperature change $\theta^{(\beta \gamma)}$ within any subcell $(\beta \gamma)$, as produced by the application
of an external electric field, $E_0$, can be determined from Eqs. (\ref{T3}) and (\ref{T4}).
The derived numerical integration scheme appears to be stable (in terms of the recursion mapping encompassed in the outlined overall matrix equation comprised of Eqs. (\ref{T10}-\ref{T12})).

\section{The Mechanical Field}
\label{Sect6}

The method of solution of the mechanical problem is also similar to the HOT approach that has been presented in chapter 11 of \cite{AAB}, which deals with functionally graded materials.
Presently, however, the incorporation of the ponderomotive force and the temperature field need to be dealt with.
Thus, the solution of the mechanical problem is based on satisfying the governing equations, continuity and boundary conditions in the average (integral) sense. 
This implies that the equilibrium equation is solved in strong form (i.e., not by numerical variational minimization)
yet in the volume-integral sense. Furthermore, the four three-component force vectors acting on the edges of a rectangular two-dimensional subcell,
are balanced by the three-component body force vector acting on the volume of the subcell. This results in three equations in each subcell.
In addition, the continuity conditions are enforced in the integral sense, that is forces (normalized by edge areas) acting on the edges of adjacent subcells are set equal,
which gives six more equations for each subcell. Finally, the displacements continuity is enforced. In principle, displacements could be made continuous
at the centers of the edges, that is locally, just as was done in the case of the temperatures in the thermal problem.
However, unlike the temperature, which is a purely local, intensive state variable in the material, the displacement is in fact a spatially integrable quantity.
Consequently, displacement continuity is enforced here for the center-of-edge displacements, that is for the integrals of the displacement field along the edge,
normalized by the edge length, rather than for the local displacement of the centers of the edges. This gives another six equations, leading to a total of fifteen for a subcell.
This calls for fifteen subcell variables used to interpolate the displacement vector field within the subcell, as a quadratic polynomial in two variables
(suitable for the plane deformation assumption), without the bilinear term. 

The displacement components within subcell $(\beta \gamma)$ are, therefore, expanded into second-order polynomials, as follows:
\begin{eqnarray} \label{S1}
u^{(\beta \gamma)}_1 &=& W^{(\beta \gamma)}_{1(00)} +
{y}^{(\beta )}_{2}  W^{(\beta \gamma)}_{1(10)}+
{y}^{(\gamma)}_{3}  W^{(\beta \gamma)}_{1(01)}  \nonumber  \\
 &+&  \frac{1}{2} \left[3 \left({y}^{(\beta )}_2\right)^2 -\frac{h^2_{\beta }}{4} \right] W^{(\beta \gamma)}_{1(20)}
  +   \frac{1}{2} \left[3 \left({y}^{(\gamma)}_3\right)^2 -\frac{l^2_{\gamma}}{4} \right] W^{(\beta \gamma)}_{1(02)}
\end{eqnarray}
\begin{eqnarray} \label{S2}
u^{(\beta \gamma)}_2 &=& W^{(\beta \gamma)}_{2(00)} +
{y}^{(\beta )}_{2}  W^{(\beta \gamma)}_{2(10)}+
{y}^{(\gamma)}_{3}  W^{(\beta \gamma)}_{2(01)}  \nonumber  \\
 &+&  \frac{1}{2} \left[3 \left({y}^{(\beta )}_2\right)^2 -\frac{h^2_{\beta }}{4} \right] W^{(\beta \gamma)}_{2(20)}
  +   \frac{1}{2} \left[3 \left({y}^{(\gamma)}_3\right)^2 -\frac{l^2_{\gamma}}{4} \right]W^{(\beta \gamma)}_{2(02)}
\end{eqnarray}
\begin{eqnarray} \label{S3}
u^{(\beta \gamma)}_3 &=& W^{(\beta \gamma)}_{3(00)} +
{y}^{(\beta )}_{2}  W^{(\beta \gamma)}_{3(10)}+
{y}^{(\gamma)}_{3}  W^{(\beta \gamma)}_{3(01)}  \nonumber  \\
 &+&  \frac{1}{2} \left[3 \left({y}^{(\beta )}_2\right)^2 -\frac{h^2_{\beta }}{4} \right] W^{(\beta \gamma)}_{3(20)}
  +   \frac{1}{2} \left[3 \left({y}^{(\gamma)}_3\right)^2 -\frac{l^2_{\gamma}}{4} \right]W^{(\beta \gamma)}_{3(02)}
\end{eqnarray}
The resulting strain components in subcell $(\beta \gamma)$ are given by:
\begin{eqnarray} \label{S4}
\epsilon^{(\beta \gamma)}_{11} &=& 0       \nonumber \\             
\epsilon^{(\beta \gamma)}_{22} &=& W^{(\beta \gamma)}_{2(10)} + 3 {y}^{(\beta )}_2 W^{(\beta \gamma)}_{2(20)}                     \nonumber \\
\epsilon^{(\beta \gamma)}_{33} &=& W^{(\beta \gamma)}_{3(01)} + 3 {y}^{(\gamma)}_3 W^{(\beta \gamma)}_{3(02)}                     \nonumber \\
\epsilon^{(\beta \gamma)}_{23} &=& \frac{1}{2} \left(W^{(\beta \gamma)}_{2(01)} + 3 {y}^{(\gamma)}_3 W^{(\beta \gamma)}_{2(02)}
                                +W^{(\beta \gamma)}_{3(10)} + 3 {y}^{(\beta )}_2 W^{(\beta \gamma)}_{3(20)} \right)                 \nonumber \\
\epsilon^{(\beta \gamma)}_{13} &=& \frac{1}{2} \left(W^{(\beta \gamma)}_{1(01)}
                                + 3 {y}^{(\gamma)}_3 W^{(\beta \gamma)}_{1(02)} \right)                                          \nonumber \\
\epsilon^{(\beta \gamma)}_{12} &=& \frac{1}{2} \left(W^{(\beta \gamma)}_{1(10)}
                                + 3 {y}^{(\beta )}_2 W^{(\beta \gamma)}_{1(20)} \right)
\end{eqnarray}

By averaging the equilibrium equations (\ref{E3}) over the subcell volume (area in the two-dimensional case considered here),
one obtains, in conjunction with Eqs. (\ref{S4}) and (\ref{E7}), the following relations:
\begin{eqnarray} \label{S5}
C^{(\beta \gamma)}_{66} W^{(\beta \gamma)}_{1(20)} + C^{(\beta \gamma)}_{55} W^{(\beta \gamma)}_{1(02)}                                                         &=& 0  \nonumber \\ 
C^{(\beta \gamma)}_{22} W^{(\beta \gamma)}_{2(20)} + C^{(\beta \gamma)}_{44} W^{(\beta \gamma)}_{2(02)} + \bar F^{(\beta \gamma)}_2 - \bar G^{(\beta \gamma)}_2 &=& 0  \nonumber \\ 
C^{(\beta \gamma)}_{44} W^{(\beta \gamma)}_{3(20)} + C^{(\beta \gamma)}_{33} W^{(\beta \gamma)}_{2(02)} + \bar F^{(\beta \gamma)}_3 - \bar G^{(\beta \gamma)}_3 &=& 0   
\end{eqnarray}
where the volume-average (area-average in the two-dimensional case) of the ponderomotive force-to-unit-volume, used in Eq. (\ref{S5}), can be expressed as: 
\begin{eqnarray} \label{S6}
\bar F^{(\beta \gamma)}_i = \frac{1}{h_{\beta} l_{\gamma}} \int^{h_{\beta}/2}_{-h_{\beta}/2} \int^{l_{\gamma}/2}_{-l_{\gamma}/2} F^{(\beta \gamma)}_i (y^{(\beta)}_2, y^{(\gamma)}_3)
dy^{(\beta)}_2 dy^{(\gamma)}_3, \ \ \ \ i=2,3  
\end{eqnarray}
and the volume (area) average of the principle components of the thermal stress involving the temperature gradient (resulting from the divergence of the thermal stress tensor,
which is spherical in the thermally isotropic case considered here) is given by:
\begin{eqnarray} \label{S7}
\bar G^{(\beta \gamma)}_2 &=& \frac{\Gamma^{(\beta \gamma)}_2}{h_{\beta} l_{\gamma}} \int^{h_{\beta}/2}_{-h_{\beta}/2} \int^{l_{\gamma}/2}_{-l_{\gamma}/2}
\frac{\partial \theta^{(\beta \gamma)} (y^{(\beta)}_2, y^{(\gamma)}_3)}{\partial y^{(\beta)}_2} dy^{(\beta)}_2 dy^{(\gamma)}_3                 \nonumber \\
\bar G^{(\beta \gamma)}_3 &=& \frac{\Gamma^{(\beta \gamma)}_3}{h_{\beta} l_{\gamma}} \int^{h_{\beta}/2}_{-h_{\beta}/2} \int^{l_{\gamma}/2}_{-l_{\gamma}/2}
\frac{\partial \theta^{(\beta \gamma)} (y^{(\beta)}_3, y^{(\gamma)}_3)}{\partial y^{(\gamma)}_3} dy^{(\beta)}_2 dy^{(\gamma)}_3  
\end{eqnarray}

By utilizing the expressions given for the ponderomotive force in Eqs. (\ref{M16}-\ref{M17}),
the volume average (area-average here) of $\bar F^{(\beta \gamma)}_i$ can be readily determined.  
As for the components of the volume (area) average of the temperature gradient, these can be determined from the expression for
$\theta^{(\beta \gamma)}$ given by Eqs. (\ref{T0}), (\ref{T3}) and (\ref{T4}), assuming the expansion coefficients are solved for.  

It should be noted that instead of writing the equilibrium equations in terms of edge-averaged tractions expressed through
stress-strain relations via the coefficients expanding the displacement field, the equivalent direct volume (area) averaging
on the equilibrium equations was employed, making use of the divergence theorem applied to a subcell. 

Now, with the stress tensor given be Eq. (\ref{E7}), the line-average stress components along the subcell edges can be expressed as follows: 
\begin{eqnarray} \label{S8}
T^{\pm (\beta \gamma)}_{2i} &=& \frac {1}{l_{\gamma}} \int_{ -l_{\gamma} / 2}^{ l_{\gamma} / 2}
   {\Sigma}^{(\beta \gamma)}_{2i} \left({y}^{(\beta)}_2 = \pm \frac{ h_{\beta}} {2} \right) \ d {y}^{(\gamma)}_3 \nonumber \\
T^{\pm (\beta \gamma)}_{3i} &=& \frac {1}{h_{\beta}} \int_{ -h_{\beta} / 2}^{ h_{\beta} / 2}
   {\Sigma}^{(\beta \gamma)}_{3i} \left({y}^{(\gamma)}_3 = \pm \frac{ l_{\gamma}} {2} \right) \ d {y}^{(\beta)}_2
\end{eqnarray}

By substituting Eqs. (\ref{E7}) and (\ref{S4}) in Eq. (\ref{S8}), the following expressions are obtained:
\begin{eqnarray} \label{S9}
T^{\pm (\beta \gamma)}_{21} &=& C^{(\beta \gamma)}_{66} \left[ W^{(\beta \gamma)}_{1(10)} \pm \frac{3  h_{\beta}}{2} W^{(\beta \gamma)}_{1(20)} \right]                              \nonumber \\
T^{\pm (\beta \gamma)}_{22} &=& C^{(\beta \gamma)}_{22} \left[ W^{(\beta \gamma)}_{2(10)} \pm \frac{3  h_{\beta}}{2} W^{(\beta \gamma)}_{2(20)} \right]
    + C^{(\beta \gamma)}_{23} W^{(\beta \gamma)}_{3(01)} - \bar \Gamma^{\pm(\beta \gamma)}_2                                                                                         \nonumber \\
T^{\pm (\beta \gamma)}_{23} &=& C^{(\beta \gamma)}_{44} \left[ W^{(\beta \gamma)}_{2(01)} + W^{(\beta \gamma)}_{3(10)} \pm \frac{3  h_{\beta}}{2} W^{(\beta \gamma)}_{3(20)} \right] \nonumber \\
T^{\pm (\beta \gamma)}_{31} &=& C^{(\beta \gamma)}_{55} \left[ W^{(\beta \gamma)}_{1(01)} \pm \frac{3  l_{\gamma}}{2} W^{(\beta \gamma)}_{1(02)} \right]                             \nonumber \\
T^{\pm (\beta \gamma)}_{32} &=& C^{(\beta \gamma)}_{44} \left[ W^{(\beta \gamma)}_{2(01)} + W^{(\beta \gamma)}_{3(10)} \pm \frac{3  l_{\gamma}}{2} W^{(\beta \gamma)}_{2(02)} \right] \nonumber \\
T^{\pm (\beta \gamma)}_{33} &=& C^{(\beta \gamma)}_{23} W^{(\beta \gamma)}_{2(10)} + C^{(\beta \gamma)}_{33} \left[ W^{(\beta \gamma)}_{3(01)} \pm \frac{3 l_{\gamma}}{2} W^{(\beta \gamma)}_{3(02)} \right]
      - \bar \Gamma^{\pm(\beta \gamma)}_3                                                                                             
\end{eqnarray}
where the definition of edge-averaged thermal stresses is introduced, as follows:
\begin{eqnarray} \label{S10}
\bar \Gamma^{\pm (\beta \gamma)}_{2} &=& \frac{\Gamma^{(\beta \gamma)}_2}{l_{\gamma}} \int^{l_{\gamma}/2}_{-l_{\gamma}/2} \theta^{(\beta \gamma)} \left(\pm h_{\beta }/2, y^{(\gamma)}_3 \right)
                                      dy^{(\gamma)}_3  \nonumber \\
\bar \Gamma^{\pm (\beta \gamma)}_{3} &=& \frac{\Gamma^{(\beta \gamma)}_3}{h_{\beta }} \int^{h_{\beta }/2}_{-h_{\beta }/2} \theta^{(\beta \gamma)} \left(y^{(\beta)}_2, \pm l_{\gamma} / 2 \right)    
                                      dy^{(\beta )}_2  
\end{eqnarray}

By recalling Eq. (\ref{T4}), edge-averaged thermal stresses can be obtained explicitly, as:

\begin{eqnarray} \label{S11}
\bar \Gamma^{\pm (\beta \gamma)}_{2} &=& \Gamma^{ (\beta \gamma)}_{2} \left[\frac{4}{\sqrt{2} \pi l_{\gamma}} \sinh \left( \frac{\pi l_{\gamma}}{4 h_{\beta }} \right) \hat \theta^{(\beta \gamma)}_1  
                                      +  \frac{4}{\sqrt{2} \pi h_{\beta }} \cosh \left( \frac{\pi h_{\beta }}{4 l_{\gamma}} \right) \hat \theta^{(\beta \gamma)}_4    \right.         \nonumber  \\ 
                                     &-& \left. \frac{E^2_0}{48 h_{\beta}} \left( 3 h_{\beta}^2 + l_{\gamma}^2 \right) \frac{\sigma^{(\beta \gamma)}}{\kappa^{(\beta \gamma)}} 
                                   \pm \frac{4}{\sqrt{2} \pi h_{\beta }} \sinh \left( \frac{\pi h_{\beta }}{4 l_{\gamma}} \right) \hat \theta^{(\beta \gamma)}_3  \right]            \nonumber \\
\bar \Gamma^{\pm (\beta \gamma)}_{3} &=& \Gamma^{ (\beta \gamma)}_{3} \left[ \frac{4}{\sqrt{2} \pi l_{\gamma}} \cosh \left( \frac{\pi l_{\gamma}}{4 h_{\beta }} \right) \hat \theta^{(\beta \gamma)}_1  
                                      +  \frac{4}{\sqrt{2} \pi h_{\beta }} \sinh \left( \frac{\pi h_{\beta }}{4 l_{\gamma}} \right) \hat \theta^{(\beta \gamma)}_4    \right.         \nonumber \\
                                     &-&\left.  \frac{E^2_0}{48 l_{\gamma}} \left( h_{\beta}^2 + 3 l_{\gamma}^2 \right) \frac{\sigma^{(\beta \gamma)}}{\kappa^{(\beta \gamma)}} 
                                   \pm \frac{4}{\sqrt{2} \pi l_{\gamma}} \sinh \left( \frac{\pi l_{\gamma }}{4 h_{\beta}} \right) \hat \theta^{(\beta \gamma)}_2  \right]          
\end{eqnarray}

The line-averaged displacement components along the subcell edges are similarly defined as:

\begin{eqnarray} \label{S12}
U^{\pm (\beta \gamma)}_{2i} &=& \frac {1}{l_{\gamma}} \int_{ -l_{\gamma} / 2}^{ l_{\gamma} / 2}
   {u}^{(\beta \gamma)}_{2} \left({y}^{(\beta)}_2 = \pm \frac{ h_{\beta}} {2} \right) \ d {y}^{(\gamma)}_3 \nonumber \\
U^{\pm (\beta \gamma)}_{3i} &=& \frac {1}{h_{\beta}} \int_{ -h_{\beta} / 2}^{ h_{\beta} / 2}
   {u}^{(\beta \gamma)}_{3} \left({y}^{(\gamma)}_3 = \pm \frac{ l_{\gamma}} {2} \right) \ d {y}^{(\beta)}_2
\end{eqnarray}

Substituting the displacement expansions given in Eqs.(\ref{S1})-(\ref{S3}) into Eq. (\ref{S12}) gives:

\begin{eqnarray} \label{S13}
U^{\pm (\beta \gamma)}_{2i} &=& W^{(\beta \gamma)}_{i(00)} \pm \frac{h_{\beta }}{2} W^{(\beta \gamma)}_{i(10)} + \frac{h^2_{\beta }}{4} W^{(\beta \gamma)}_{i(20)} \nonumber \\
U^{\pm (\beta \gamma)}_{3i} &=& W^{(\beta \gamma)}_{i(00)} \pm \frac{l_{\gamma}}{2} W^{(\beta \gamma)}_{i(01)} + \frac{l^2_{\gamma}}{4} W^{(\beta \gamma)}_{i(02)}
\end{eqnarray}

Next, static condensation of the equilibrium equations with three of the coefficients of the expansion of the displacement field vector is performed.
The remaining twelve expansion coefficients are consequently related to the twelve edge-averaged displacements, which are then tied to the
twelve displacement/traction continuity conditions of each subcell.

First, manipulation of pairs in Eqs. (\ref{S13}), results in:

\begin{eqnarray} \label{S14}
W^{(\beta \gamma)}_{i(10)} &=& \frac{1}{h_{\beta}}  \left( U^{+(\beta \gamma)}_{2i} - U^{-(\beta \gamma)}_{2i} \right)  \nonumber \\  
W^{(\beta \gamma)}_{i(01)} &=& \frac{1}{l_{\gamma}} \left( U^{+(\beta \gamma)}_{3i} - U^{-(\beta \gamma)}_{3i} \right)  
\end{eqnarray}
and
\begin{eqnarray} \label{S15}
W^{(\beta \gamma)}_{i(20)} &=& \frac{2}{h^2_{\beta}}  \left( U^{+(\beta \gamma)}_{2i} + U^{-(\beta \gamma)}_{2i} \right) - \frac{4}{h^2_{\beta}}  W^{(\beta \gamma)}_{i(00)}  \nonumber \\
W^{(\beta \gamma)}_{i(02)} &=& \frac{2}{l^2_{\gamma}} \left( U^{+(\beta \gamma)}_{3i} + U^{-(\beta \gamma)}_{3i} \right) - \frac{4}{l^2_{\gamma}} W^{(\beta \gamma)}_{i(00)}  
\end{eqnarray}

Next, substituting $W^{(\beta \gamma)}_{i(20)}$ and $W^{(\beta \gamma)}_{i(02)}$ into Eq. (\ref{S5}) yields the following expressions for $W^{(\beta \gamma)}_{i(00)}$: 

\begin{eqnarray} \label{S16}
W^{(\beta \gamma)}_{1(00)} &=& \frac{ C^{(\beta \gamma)}_{55} h^2_{\beta } \left( U^{+(\beta \gamma)}_{31} + U^{-(\beta \gamma)}_{31} \right)
                                   +  C^{(\beta \gamma)}_{66} l^2_{\gamma} \left( U^{+(\beta \gamma)}_{21} + U^{-(\beta \gamma)}_{21} \right)}
                                      {2 \left( C^{(\beta \gamma)}_{55} h^2_{\beta} + C^{(\beta \gamma)}_{66} l^2_{\gamma} \right) }                     \nonumber \\ 
W^{(\beta \gamma)}_{2(00)} &=& \frac{h^2_{\beta} l^2_{\gamma} \left(\bar F^{(\beta \gamma)}_2 - \bar G^{(\beta \gamma)}_2 \right) 
                                   +2 C^{(\beta \gamma)}_{44} h^2_{\beta } \left( U^{+(\beta \gamma)}_{32} + U^{-(\beta \gamma)}_{32} \right)
                                   +2 C^{(\beta \gamma)}_{22} l^2_{\gamma} \left( U^{+(\beta \gamma)}_{22} + U^{-(\beta \gamma)}_{22} \right)}
                                      {4 \left( C^{(\beta \gamma)}_{44} h^2_{\beta} + C^{(\beta \gamma)}_{22} l^2_{\gamma} \right) }                     \nonumber \\ 
W^{(\beta \gamma)}_{3(00)} &=& \frac{h^2_{\beta} l^2_{\gamma} \left(\bar F^{(\beta \gamma)}_3 - \bar G^{(\beta \gamma)}_3 \right) 
                                   +2 C^{(\beta \gamma)}_{33} h^2_{\beta } \left( U^{+(\beta \gamma)}_{33} + U^{-(\beta \gamma)}_{33} \right)
                                   +2 C^{(\beta \gamma)}_{44} l^2_{\gamma} \left( U^{+(\beta \gamma)}_{23} + U^{-(\beta \gamma)}_{23} \right)}
                                      {4 \left( C^{(\beta \gamma)}_{33} h^2_{\beta} + C^{(\beta \gamma)}_{44} l^2_{\gamma} \right) }                  \nonumber \\                   
\end{eqnarray}

By substituting the expressions for $W^{(\beta \gamma)}_{i(mn)}$ given in Eq. (\ref{S14})-(\ref{S16}) into the edge-averaged traction components in Eq. (\ref{S9}),
the following matrix equation is obtained for subcell $(\beta \gamma)$:

\begin{eqnarray} \label{S17}
\left\{ \begin{array}{c}
T^{+}_{21}   \\
T^{-}_{21}   \\
T^{+}_{22}   \\
T^{-}_{22}   \\
T^{+}_{23}   \\
T^{-}_{23}   \\
T^{+}_{31}   \\
T^{-}_{31}   \\
T^{+}_{32}   \\
T^{-}_{32}   \\
T^{+}_{33}   \\
T^{-}_{33}   \end{array} \right\}^{(\beta \gamma)}
&=&  {\bm K}^{(\beta \gamma)}
\left\{ \begin{array}{c}
U^{+}_{21}   \\
U^{-}_{21}   \\
U^{+}_{22}   \\
U^{-}_{22}   \\
U^{+}_{23}   \\
U^{-}_{23}   \\
U^{+}_{31}   \\
U^{-}_{31}   \\
U^{+}_{32}   \\
U^{-}_{32}   \\
U^{+}_{33}   \\
U^{-}_{33}   \end{array} \right\}^{(\beta \gamma)}
+ \left\{ \begin{array}{c}
S^{+}_{21} \\
S^{-}_{21} \\
S^{+}_{22} \\
S^{-}_{22} \\
S^{+}_{23} \\
S^{-}_{23} \\
S^{+}_{31} \\
S^{-}_{31} \\
S^{+}_{32} \\
S^{-}_{32} \\
S^{+}_{33} \\
S^{-}_{33} \end{array} \right\}^{(\beta \gamma)}
\end{eqnarray}
where ${\bm K}^{(\beta \gamma)}$ is a $12 \times 12$ matrix, the elements of which depend on the dimensions of subcell $(\beta \gamma)$
and the material properties in this subcell, and where the ponderomotive force and temperature distributions' contributions are given by:

\begin{eqnarray} \label{S18}
S^{\pm (\beta \gamma)}_{21} &=& 0                                            \nonumber \\
S^{\pm (\beta \gamma)}_{22} &=& \pm \frac{3}{2}\frac{C^{(\beta \gamma)}_{22} h_{\beta} l^2_{\gamma}}{C^{(\beta \gamma)}_{22} l^2_{\gamma} + C^{(\beta \gamma)}_{44} h^2_{\beta}}
                 \left(\bar G^{(\beta \gamma)}_2 - \bar F^{(\beta \gamma)}_2 \right) - \bar\Gamma^{\pm (\beta \gamma)}_2   \nonumber \\
S^{\pm (\beta \gamma)}_{23} &=& \pm  \frac{3}{2}\frac{ C^{(\beta \gamma)}_{44} h_{\beta} l^2_{\gamma}}{C^{(\beta \gamma)}_{44} l^2_{\gamma} + C^{(\beta \gamma)}_{33} h^2_{\beta}}
                 \left(\bar G^{(\beta \gamma)}_3 - \bar F^{(\beta \gamma)}_3 \right)   \nonumber \\
S^{\pm (\beta \gamma)}_{31} &=& 0                                            \nonumber \\
S^{\pm (\beta \gamma)}_{32} &=& \pm \frac{3}{2} \frac{C^{(\beta \gamma)}_{44} h^2_{\beta} l_{\gamma}}{C^{(\beta \gamma)}_{22} l^2_{\gamma} + C^{(\beta \gamma)}_{44} h^2_{\beta}}
                 \left(\bar G^{(\beta \gamma)}_2 - \bar F^{(\beta \gamma)}_2 \right)   \nonumber \\
S^{\pm (\beta \gamma)}_{33} &=& \pm  \frac{3}{2}\frac{ C^{(\beta \gamma)}_{33} h^2_{\beta} l_{\gamma}}{C^{(\beta \gamma)}_{44} l^2_{\gamma} + C^{(\beta \gamma)}_{33} h^2_{\beta}}
                 \left(\bar G^{(\beta \gamma)}_3 - \bar F^{(\beta \gamma)}_3 \right) - \bar\Gamma^{\pm (\beta \gamma)}_3   
\end{eqnarray}

Evidently, the total number of unknowns $U^{\pm (\beta \gamma)}_{2i}, U^{\pm (\beta \gamma)}_{3i}$ in all subcells is $12N_{\beta} N_{\gamma}$. 

Now, continuity of tractions (forces) between subcells is expressed in the following equations:
\begin{eqnarray} \label{S19}
T^{+(\beta \gamma)}_{2i}  = T^{-(\beta+1, \gamma)}_{2i},   \ \ \ \ \ i=1,2,3, \ \ \ \ \ \ \beta = 1, ...,N_{\beta}-1, \ \ \ \ \ \gamma=1,...,N_{\gamma}    \nonumber \\
T^{+(\beta \gamma)}_{3i}  = T^{-(\beta  , \gamma+1)}_{3i}, \ \ \ \ \ i=1,2,3, \ \ \ \ \ \ \beta = 1, ...,N_{\beta},   \ \ \ \ \ \gamma=1,...,N_{\gamma}-1   
\end{eqnarray}
whereas continuity of edge-averaged displacements reads:
\begin{eqnarray} \label{S20}
U^{+(\beta \gamma)}_{2i}  = U^{-(\beta+1, \gamma)}_{2i},   \ \ \ \ \ i=1,2,3, \ \ \ \ \ \ \beta = 1, ...,N_{\beta}-1, \ \ \ \ \ \gamma=1,...,N_{\gamma}    \nonumber \\
U^{+(\beta \gamma)}_{3i}  = U^{-(\beta  , \gamma+1)}_{3i}, \ \ \ \ \ i=1,2,3, \ \ \ \ \ \ \beta = 1, ...,N_{\beta},   \ \ \ \ \ \gamma=1,...,N_{\gamma}-1   
\end{eqnarray}

These conditions form $12N_{\beta} N_{\gamma}-6 N_{\beta} - 6 N_{\gamma}$ equations.

Finally, assuming that the external boundaries of the composite specimen are traction-free, the global boundary conditions add the following set of equations:  
\begin{eqnarray} \label{S21}
\Sigma^{(\beta \gamma)}_{2i} &=& 0, \ \ \ \ x_2 = \pm H/2 \nonumber \\
\Sigma^{(\beta \gamma)}_{3i} &=& 0, \ \ \ \ x_3 = \pm L/2 
\end{eqnarray}
where $\beta = 1,...,N_{\beta}$ and $\gamma = 1,...,N_{\gamma}$.
These conditions form additional $6 ( N_{\beta} + N_{\gamma})$ equations, which along with the aforementioned interface continuity
conditions provide the total of required $12N_{\beta}N_{\gamma}$ equations.  
The solution of this sparse system of algebraic equations provides the variables $U^{\pm (\beta \gamma)}_{2i}$ and $U^{\pm (\beta \gamma)}_{3i}$, from which the
displacements, strains and stresses can be determined at any point in the composite.  

The final step in this procedure is proper placement of the composite specimen in space, the so-called displacement pinning, used in
problems with all-free global boundary conditions to prevent matrix singularity in the final equation system, related to rigid body motion.
In the considered case, in order to prevent rigid-body displacement, the entire edge-averaged displacement vector of the left and bottom edges of the bottom left subcell is set to zero:
\begin{eqnarray} \label{S22}
U^{- (1,1)}_{2i} =0  , \ 
U^{- (1,1)}_{3i} =0
\end{eqnarray}

Accordingly, static condensation of the overall system of equations is performed, eliminating these  six displacement-related variables, along with six equations they are involved in.

\section{Applications}
\label{Sect7}

The presented theory applies to composites with any number of fibers of arbitrary cross-section shape that should be distributed in a symmetric fashion.
The theory is next illustrated for the case of a composite of a rectangular cross-section consisting of an epoxy matrix
reinforced by two symmetrically allocated iron fibers of round cross-section, see Fig. \ref{Fig1}(c). The iron volume ratio is: $v_f$ = 0.7.
The composite is divided into $N_{\beta}$ = 100, $N_{\gamma}$ = 50 identical square subcells. The cross-sectional dimensions are macroscopic, with: H = 0.1 m and L = 0.05 m.  
The material properties of the iron fibers and the epoxy matrix are given in Table 1.  
In regard with the thermal problem, the convective boundary conditions were applied once with a Nusselt number of Nu = 1,
which corresponds to nearly pure heat conduction from the composite to the air through a boundary layer of the order of magnitude of the cross-section,
and once for a Nusselt number of Nu = 1000, corresponding to forced convection into ambient air, with the maximum typical convection coefficient, as given in \cite{Holman}.
Clearly, in practical applications, more precise knowledge of the thermal problem in the air surrounding the specimen would be beneficial.
Here, however, only the solid-mechanical problem is treated, and the convection coefficient is assumed to be known in advance. 

The volume-averages of the $x_2$ components of the ponderomotive body force-per-unit-volume in the upper ($f$ = 1) and lower ($f$ = 2) fiber are defined by:
\begin{eqnarray} \label{A1}
\bar F^{(f)}_2 = \sum^{N_{\beta}}_{\beta=1} \sum^{N_{\gamma}}_{\gamma=1} h_{\beta} l_{\gamma} {\bar F}^{(\beta \gamma)}_2, \ \ \ \ (\beta \gamma) \in f=1,2
\end{eqnarray}
where $\bar {F}^{(\beta \gamma)}_2$ is as defined in Eq. (\ref{S6}).
It should be interesting to consider the pure particle-in-vacuum Lorentz force defined in a subcell $(\beta \gamma)$ by: 
\begin{eqnarray} \label{A2}
\bar {\bm{F}}^{L(\beta \gamma)} = \frac{1}{h_{\beta} l_{\gamma}} \int^{h_{\beta}/2}_{-h_{\beta}/2} \int^{l_{\gamma}/2}_{-l_{\gamma}/2}
          [{\bm J} \times {\bm B}] (y^{(\beta)}_2, y^{(\gamma)}_3) dy^{(\beta)}_2 dy^{(\gamma)}_3  
\end{eqnarray}

This forms a counterpart of the ponderomotive force defined in Eq. (\ref{S6}).
Furthermore, the average of the second components of the Lorentz force in the upper ($f=1$) and lower ($f=2$) fiber can be defined by:
\begin{eqnarray} \label{A3}
\bar { F}^{*(f)}_2 \triangleq  \bar {{F}}^{L(f)}_2 = \sum^{N_{\beta}}_{\beta=1} \sum^{N_{\gamma}}_{\gamma=1} h_{\beta} l_{\gamma}
\bar {{F}}^{L(\beta \gamma)} _2,  \ \ \ \ (\beta \gamma) \in f=1,2
\end{eqnarray}

Following one of the perspectives of the present work, the forces defined above are used to illustrate the contribution of the magnetization-induced-force to the total ponderomotive force. 

The results of the calculations for the specific case outlined above, in terms of magnetic field intensities, magnetic flux densities,
ponderomotive force components, temperature, geometric strain tensor components, and equivalent stress  distributions, are given in
color-maps in Figs. \ref{Fig2},\ref{Fig3},\ref{Fig5},\ref{Fig6} and \ref{Fig7} for applied electric field of: $E_0$ = 0.003 V/m.
Fig. \ref{Fig4} presents the comparison of the average ponderomotive force-per volume in one fiber using once only the Lorentz part and once the total Maxwell-Minkowski model
for the considered regime, for different values of the applied electric field.
Figs. \ref{Fig8} and \ref{Fig9} exhibit the strain response of the composite for increasing values of the applied electric field in the range rendering the constitutive laws employed linear as assumed.  

\vspace{5mm} 

$\textbf{Table 1}$. Material constants for the iron fibers and epoxy matrix:
\begin{center}
\noindent
\begin{tabular}{ccccccc}
\hline
 Material                   & $E (GPa)$ & $\nu$  & $\alpha (K^{-1})$    & $\sigma (S/m)$     & $\mu_r$             & $\kappa (Wm^{-1}K^{-1})$         \\
\hline
 Iron                       &  210      & $0.25$ & $10 \times 10^{-6}$  & $1 \times 10^{7}$  &  $2 \times 10^{5}$  & 100                     \\
 Epoxy                      &  3.45     & $0.35$ & $54 \times 10^{-6}$  & 0                  &  1                  & 0.18                    \\   \hline
\end{tabular}
\end{center}
\medskip
\noindent
\small{The constant scalars $E$, $\nu$, $\alpha$, $\sigma$, $\mu_r$ and $\kappa$ denote Young's modulus, Poisson's ratio, coefficient of thermal expansion, electric conductivity,
relative permeability and thermal conductivity, respectively.}

\section{Analysis of Results}
\label{Sect8}

\begin{figure}[H]
\begin{center}
{{\includegraphics[scale = 0.6]{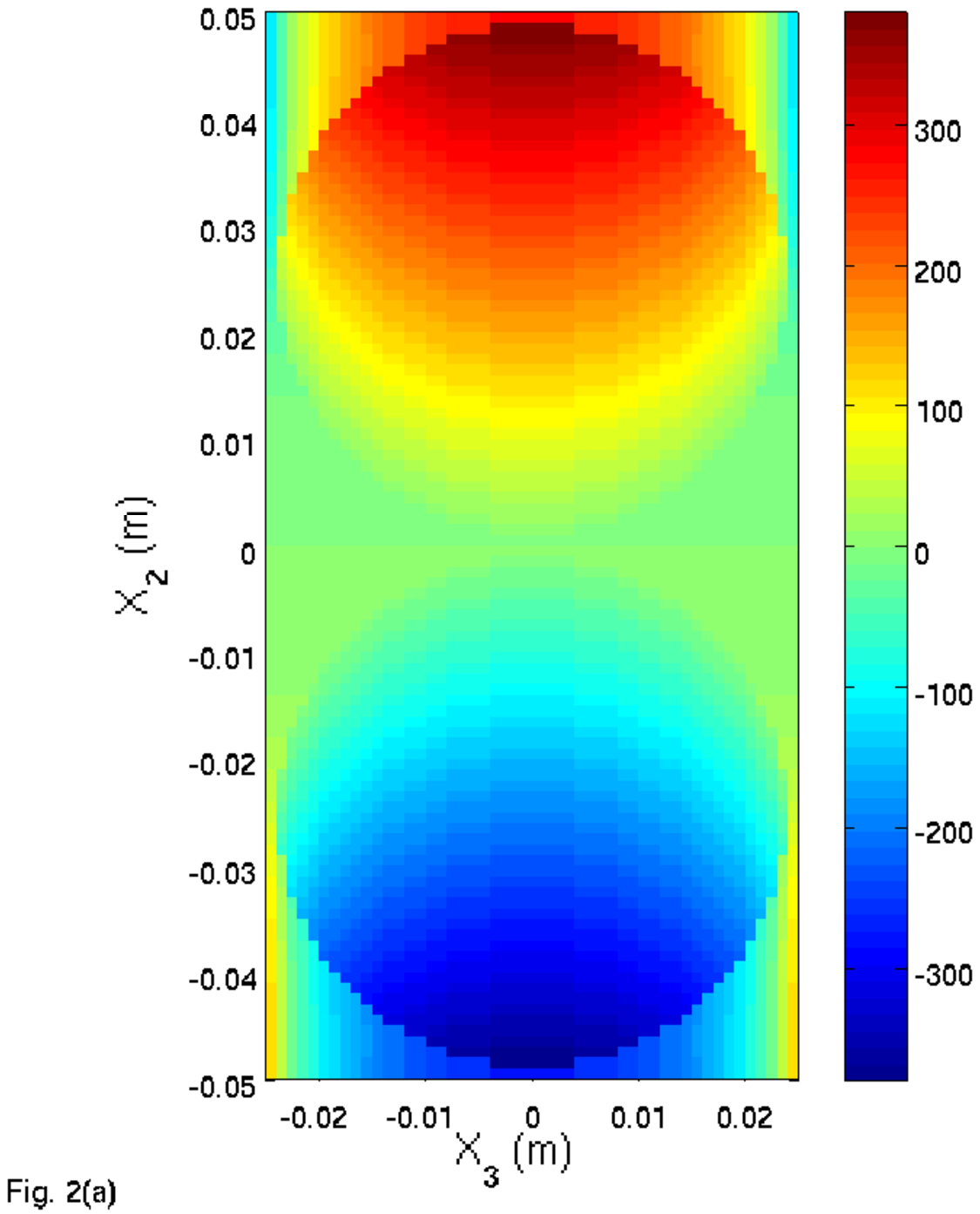}}
{\includegraphics[scale = 0.6]{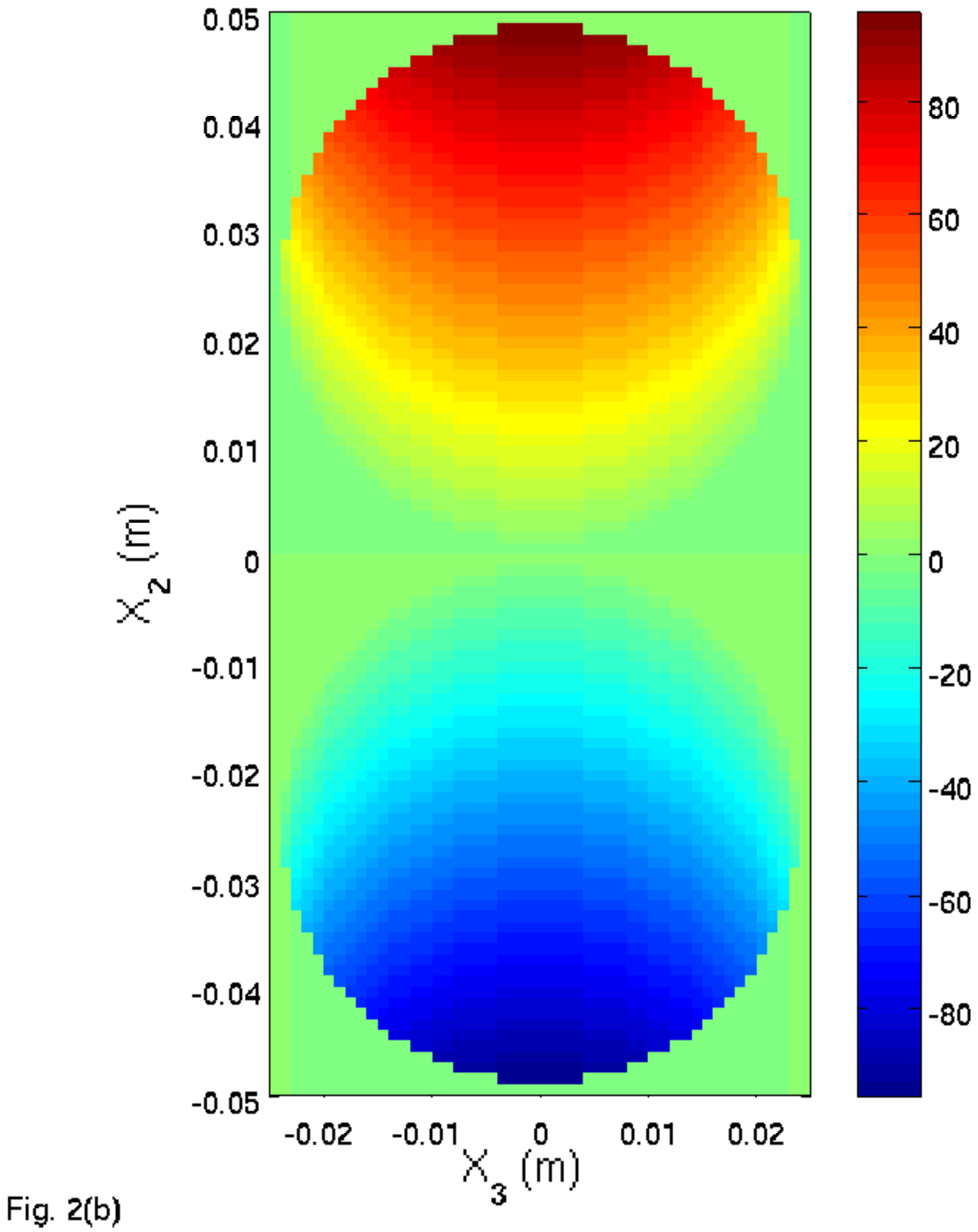}}}
\end{center}
\caption{\small The distribution of (a) $H_3$ (A/m) and, (b) $B_3$ (T) in the iron/epoxy composite (v$_f$ = 0.7) that is subjected to $E_0$ = 0.003 V/m.}
\label{Fig2}
\end{figure}

Fig. \ref{Fig2}(a) shows a color-map of the distribution of $H_3$ for $E_0$ = 0.003 V/m. One notes the continuity in the $x_2$ direction across material interface between iron and epoxy.
This is expected since $x_2$ is the direction of variation of the tangential component of the magnetic field intensity, which is just $H_3$.
Fig. \ref{Fig2}(b) shows a color-map of the distribution of $B_3$. This time the nearly identical values of $H_3$ near a material interface pointing
in the direction of $x_2$ are multiplied by permeabilities differing by orders of magnitude.
Accordingly, the appearance of tangential continuity vanishes. In addition, one may clearly see the magnetic flux density variation with $x_2$,
which shows negative values in the lower fiber, zero in the middle of the cross-section and positive values in the upper fiber,
not too different from the distribution for the case of a single fiber, where the magnetic flux density varies linearly with distance from the center.

\begin{figure}[H]
\begin{center}
{{\includegraphics[scale = 0.6]{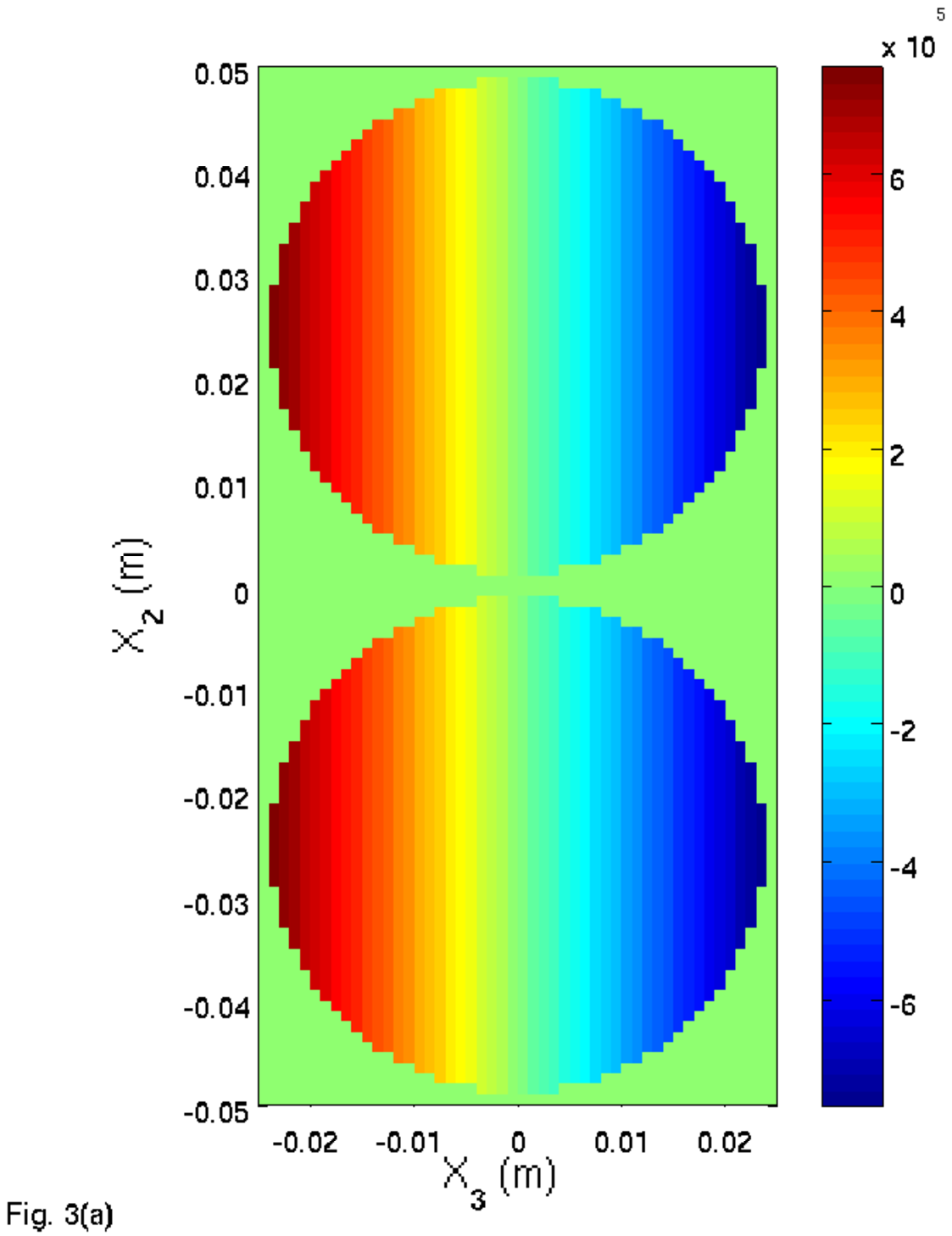}}
{\includegraphics[scale = 0.6]{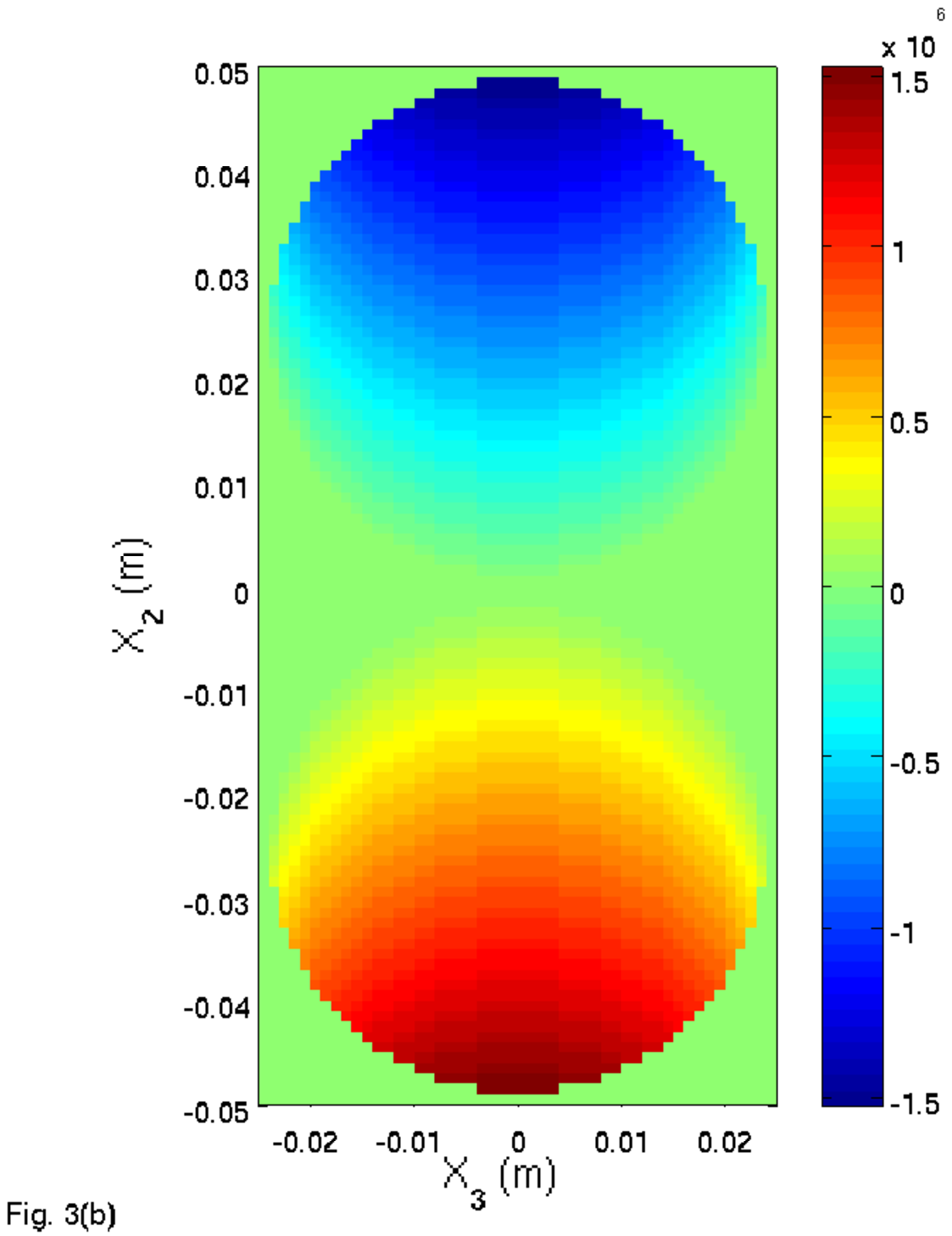}}}
\end{center}
\caption{\small The distribution of the ponderomotive force-per-unit-volume components in the iron/epoxy composite (for $v_f$ = 0.7) subjected to $E_0$ = 0.003 V/m. 
                  (a) $F_3$ (N/m$^3$), (b) $F_2$ (N/m$^3$).}
\label{Fig3}
\end{figure}

Fig. \ref{Fig3} reveals quasi-linear variation of the ponderomotive force with the coordinates, corresponding to $E_0$ = 0.003 V/m,
going through zero in the symmetry axes, implying self-contraction (material on the right experiences a negative force acting rightwards,
material on the left bears positive force rightwards, and the same goes for the vertical direction) of the widths of the fibers and
a vertical attraction force between the two fibers, as expected from experimental observation of two equal-sign-current-bearing wires,
with the additional result that the force density vanishes at the center, which is obligatory for an attraction force that has to change sign in the middle.
Also, as expected, the force vanishes in the matrix, since there is no current or magnetization there.

\begin{figure}[H]
\begin{center}
{\includegraphics[scale = 0.7]{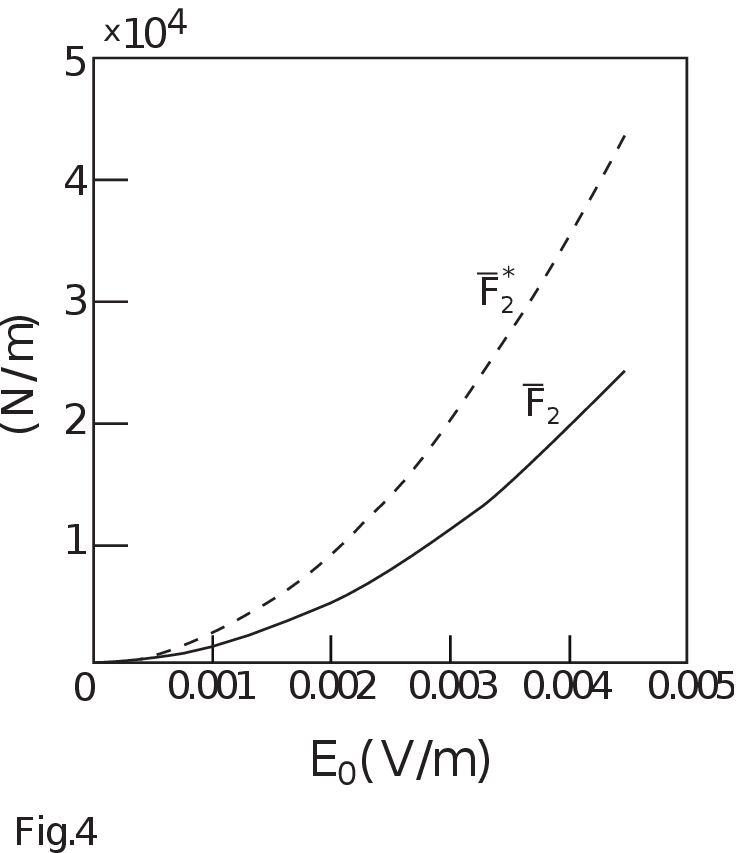}}
\end{center}
\caption{\small Comparison between the average ponderomotive force-per-unit-volume $\bar F_2$ (solid line) and $\bar F^{*}_2$ (dashed line) components in the lower fiber of Fig. \ref{Fig1}(c).}
\label{Fig4}
\end{figure}

Fig. \ref{Fig4} shows the variation of the vertical component of the ponderomotive force with applied electric field,
where the dashed line corresponds to calculation with artificially omitted magnetization (in the expression for the ponderomotive force).
One finds the expected quadratic dependence. The interesting effect here is how magnetization builds up to decrease the force experienced by the material,
thus diminishing internal magnetic energy increase caused by external work, minimizing free energy, which is expected.
The interesting part is the extent of this backlash, which reaches almost half the size of the primary phenomenon. This is of course due to the large magnetic susceptibility of iron.

\begin{figure}[H]
\begin{center}
{{\includegraphics[scale = 0.6]{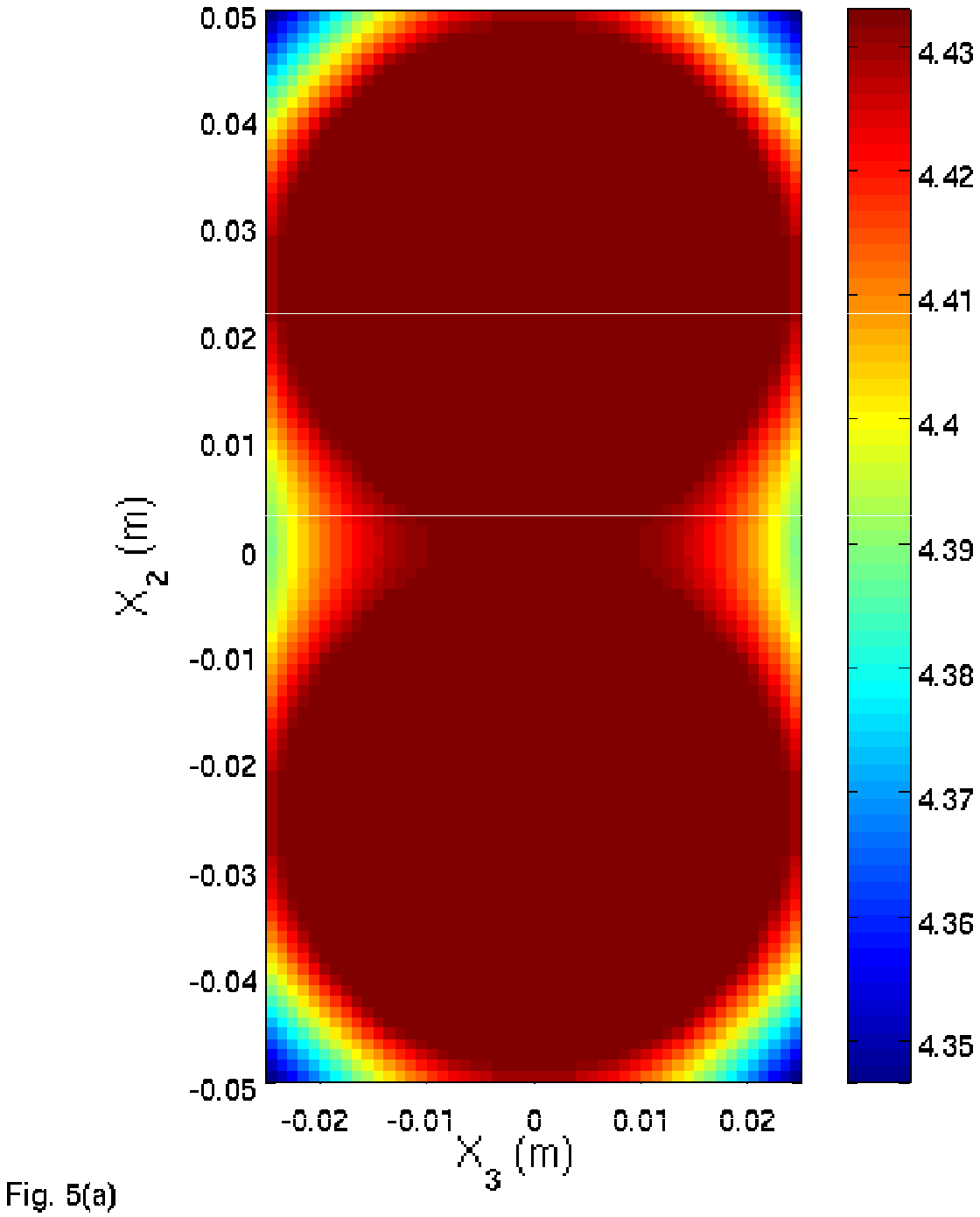}}
{\includegraphics[scale = 0.6]{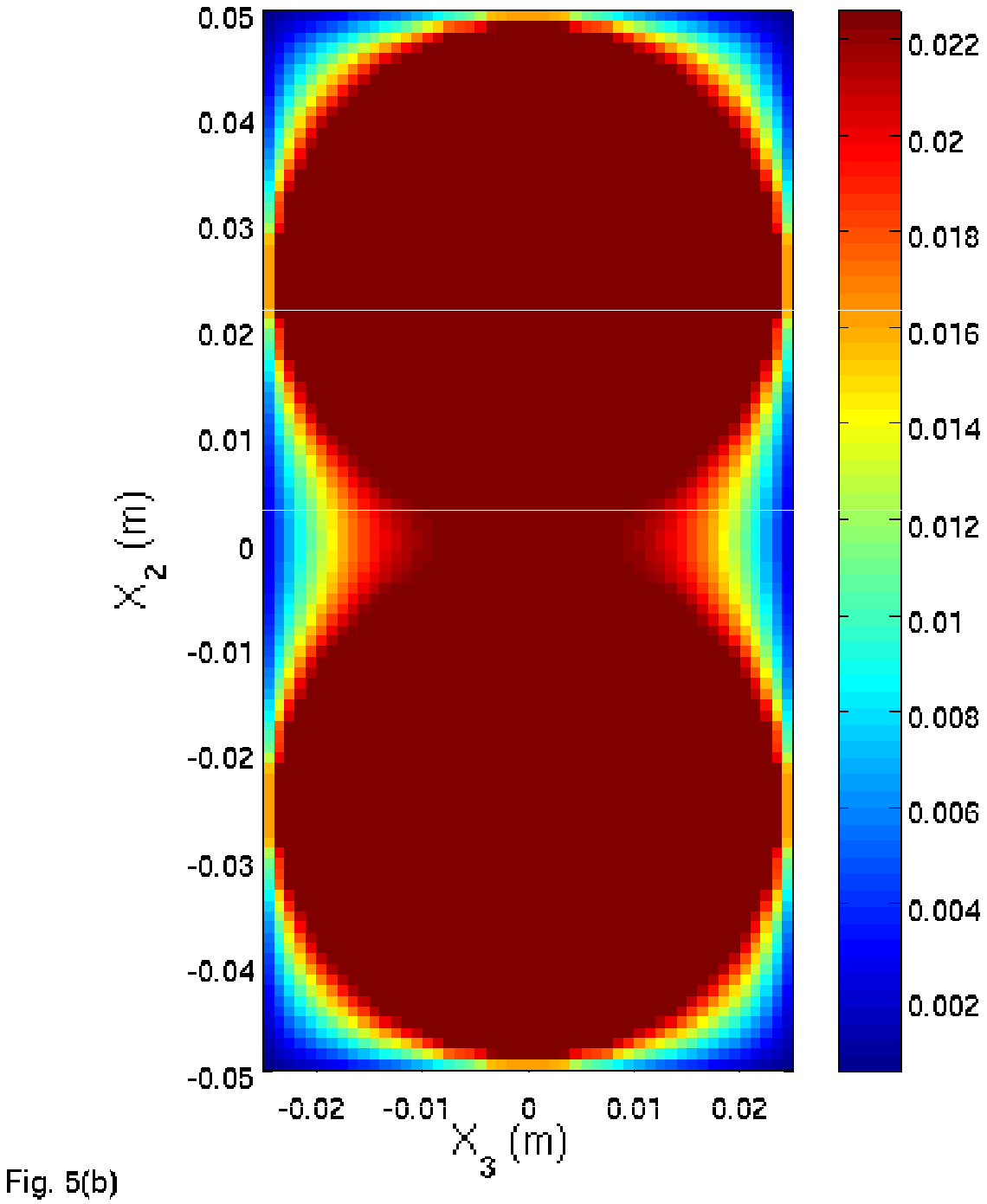}}}
\end{center}
\caption{\small The distribution of the temperature $\theta$ (K) in the iron/epoxy composite (for $v_f$ = 0.7) subjected to $E_0$ = 0.003 V/m. 
                  (a) Nu = 1, (b) Nu = 1000.}
\label{Fig5}
\end{figure}

Fig. \ref{Fig5} shows a color-map of the distribution of the temperature deviation from ambient temperature for $E_0$ = 0.003 V/m.
One can acknowledge the almost uniform distribution within the fibers, resulting from the relatively high thermal conductivity of iron,
and the relatively pronounced non-uniform temperature profile in the epoxy, corresponding to its poor thermal conduction,
the high temperature in the interface with the fiber, resulting from Joule heating, and the convection at the global boundary.
Fig. \ref{Fig5}(a) shows the case of convection by conduction, with Nu = 1, where the maximum temperature increase is about four degrees kelvin,
which is small enough to still assume constant conductivities.

Fig. \ref{Fig5}(b) shows the distributions for Nu = 1000, representing the upper limit of the typical values in forced convection by air.
The corresponding temperature deviation there is around one-fiftieth of degree kelvin, which is reasonable for
a case where generated Joule heating is very effectively removed by convection at the boundary.
Finally, the smoothness of the distribution, although not surprising for a Poisson equation with a constant inhomogeneous part,
implies that the discretization was sufficient, at least for the thermal problem.

\begin{figure}[H]
\begin{center}
{{\includegraphics[scale = 0.6]{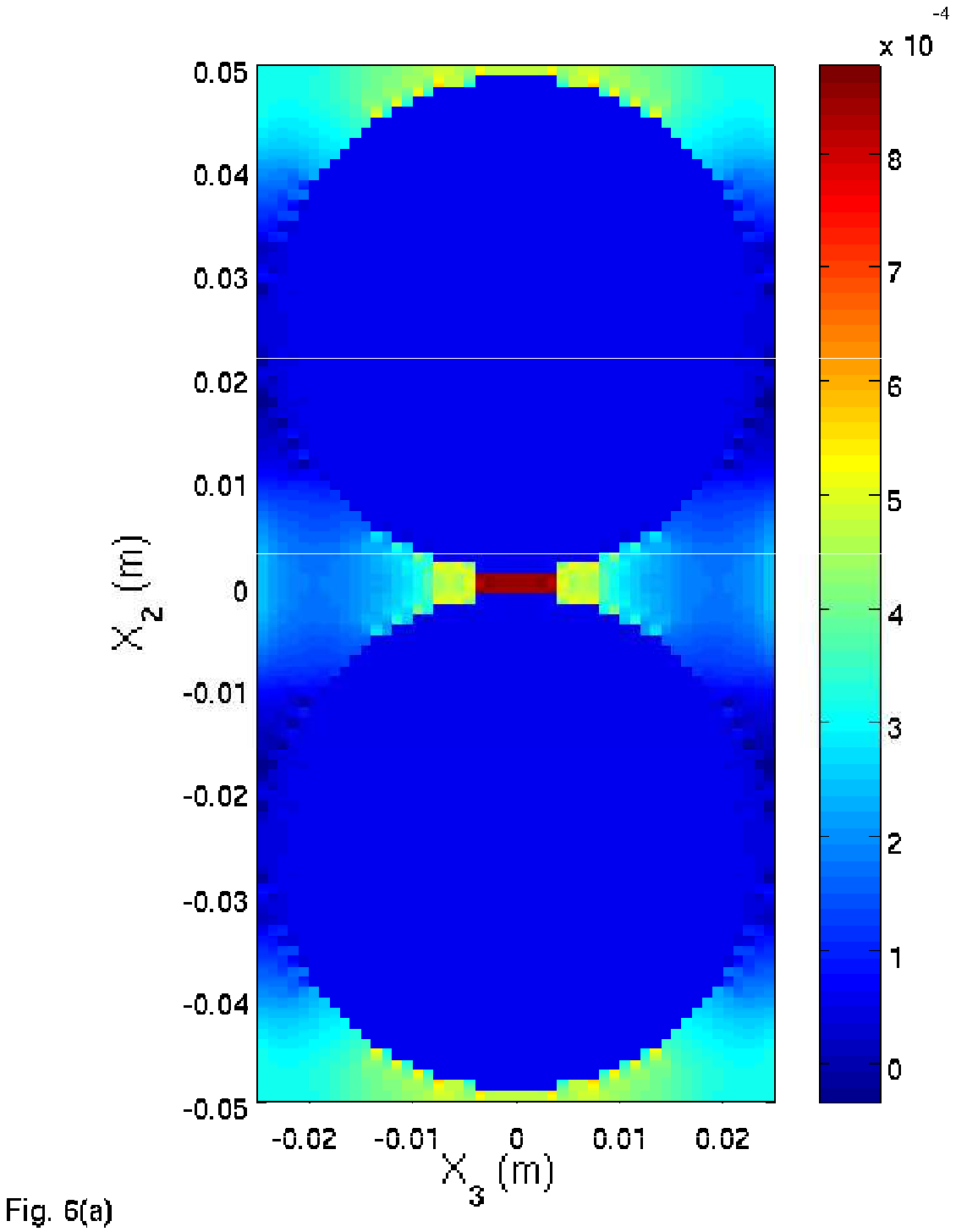}}
{\includegraphics[scale = 0.6]{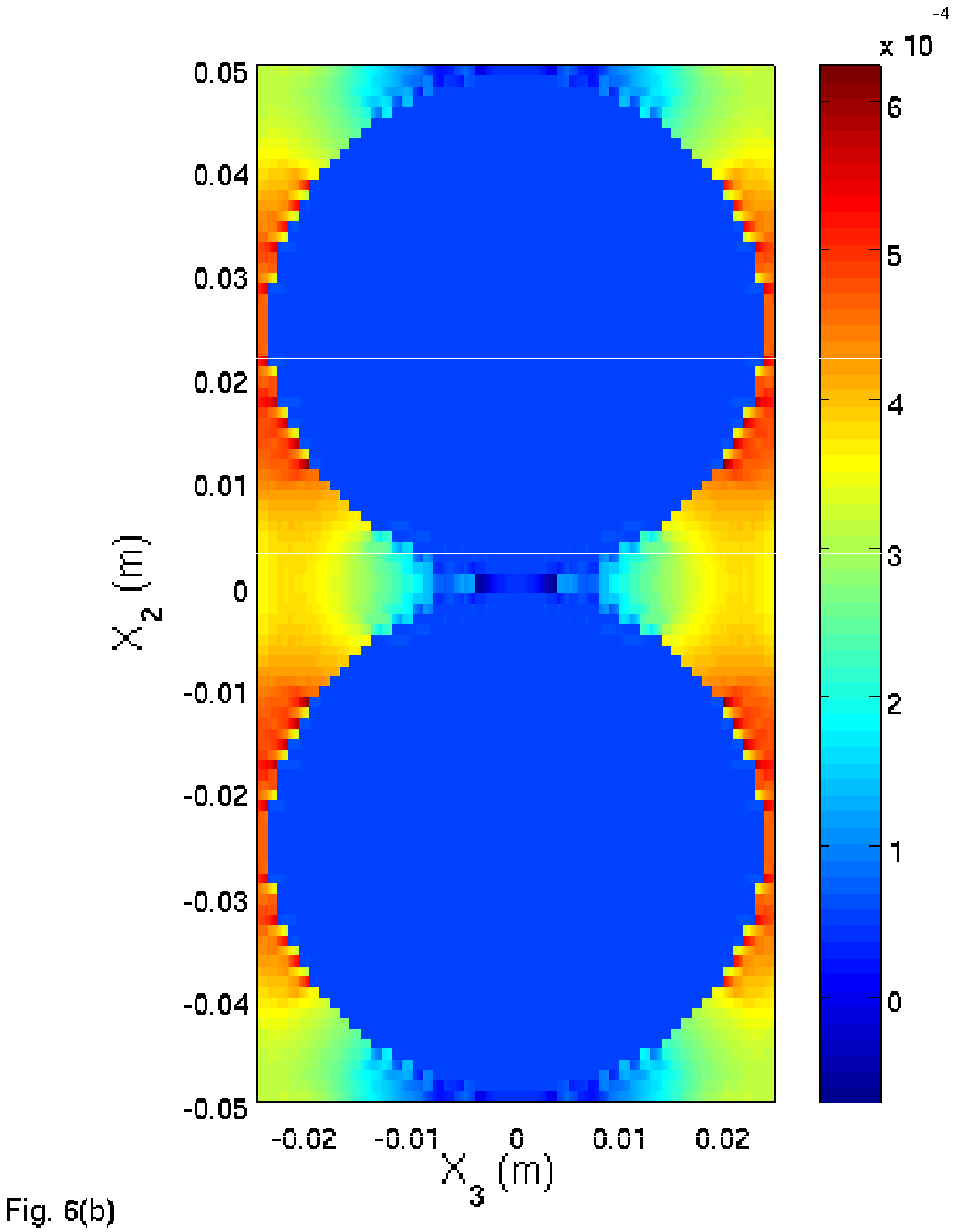}}\\
{\includegraphics[scale = 0.6]{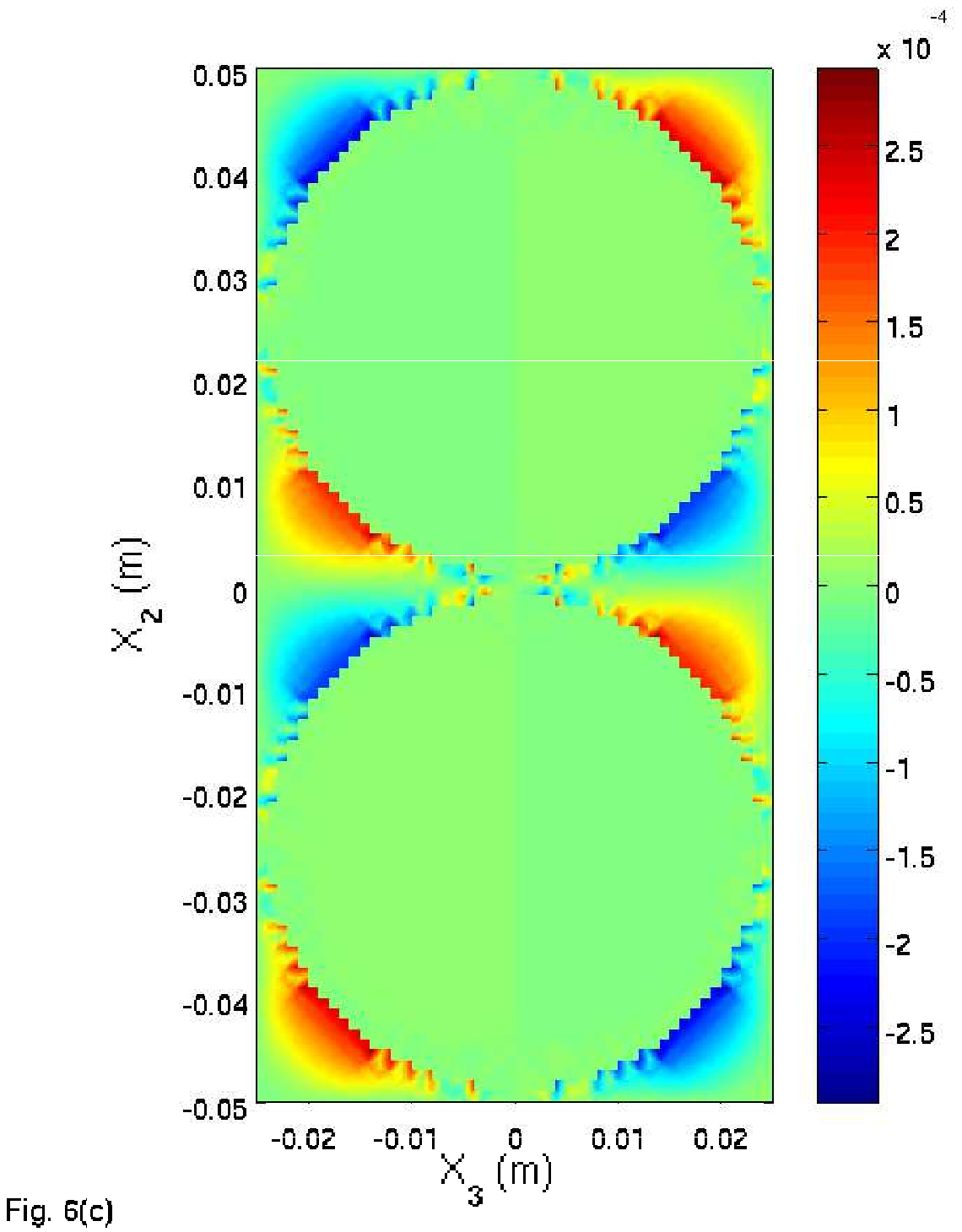}}
{\includegraphics[scale = 0.6]{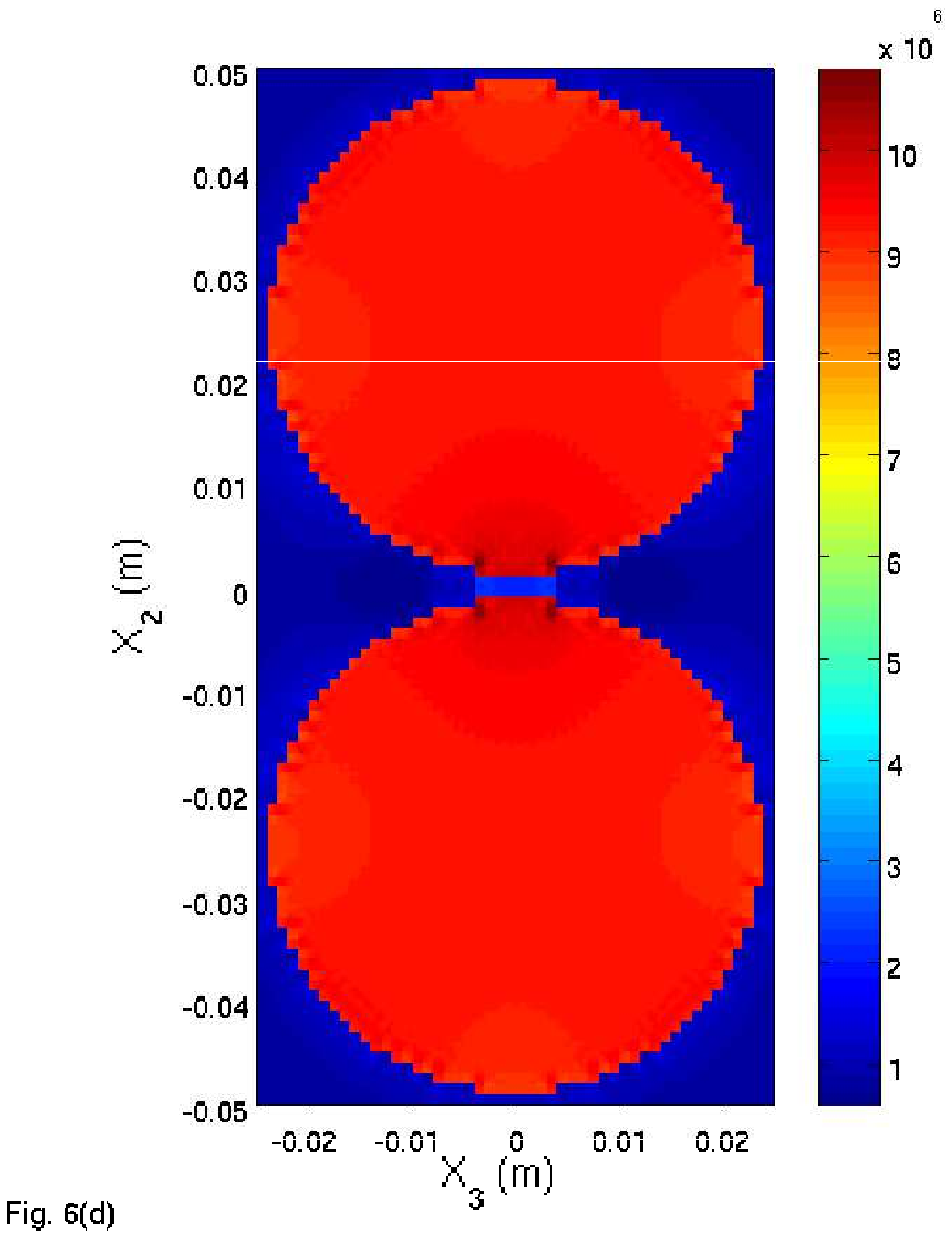}}}
\end{center}
\caption{\small The distribution of the strain components and equivalent stress in the iron/epoxy composite ($v_f$ = 0.7) subjected to $E_0$ = 0.003 V/m, with Nu = 1. 
                  (a) $\epsilon_{22}$, (b) $\epsilon_{33}$, (c) $\epsilon_{23}$, and (d) $\sigma_{eq}$ (Pa). 
                  (a) Nu = 1, (b) Nu = 1000.}
\label{Fig6}
\end{figure}

Figs. \ref{Fig6}-\ref{Fig9} are dedicated to the mechanical response of the partially conductive composite to applied electric field $E_0$ = 0.003 V/m,
which produces two competing effects: 

\begin{figure}[H]
\begin{center}
{{\includegraphics[scale = 0.6]{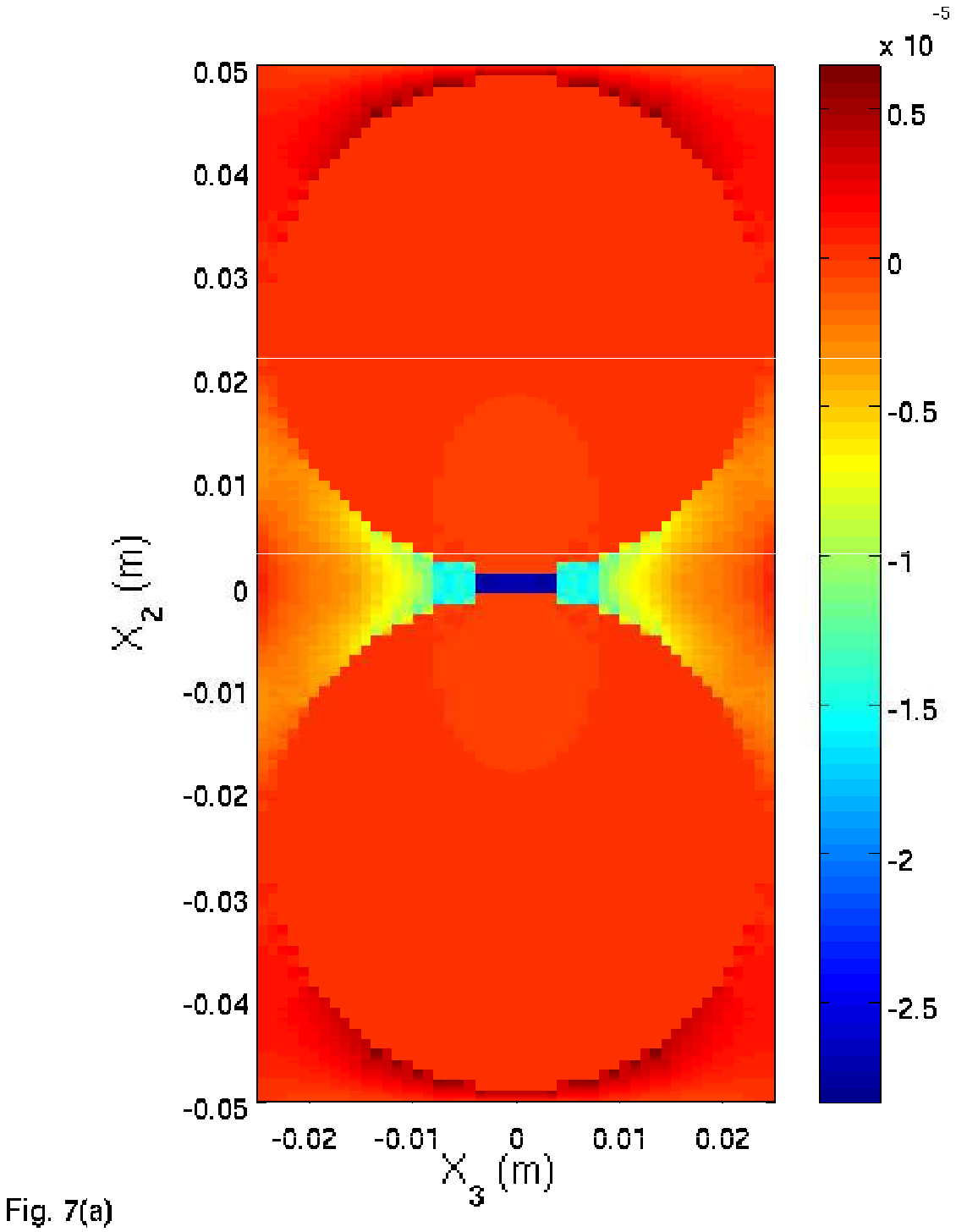}}
{\includegraphics[scale = 0.6]{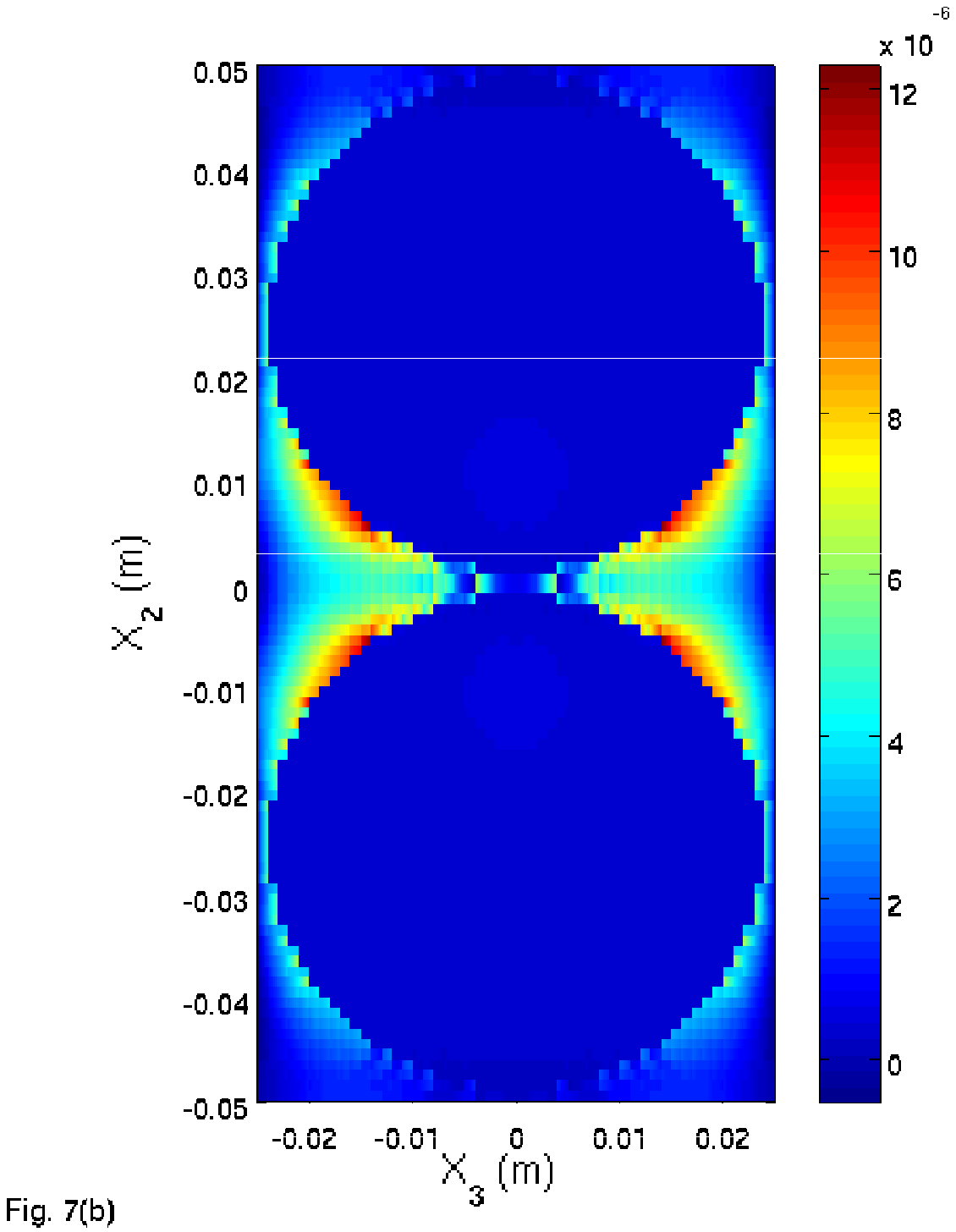}}\\
{\includegraphics[scale = 0.6]{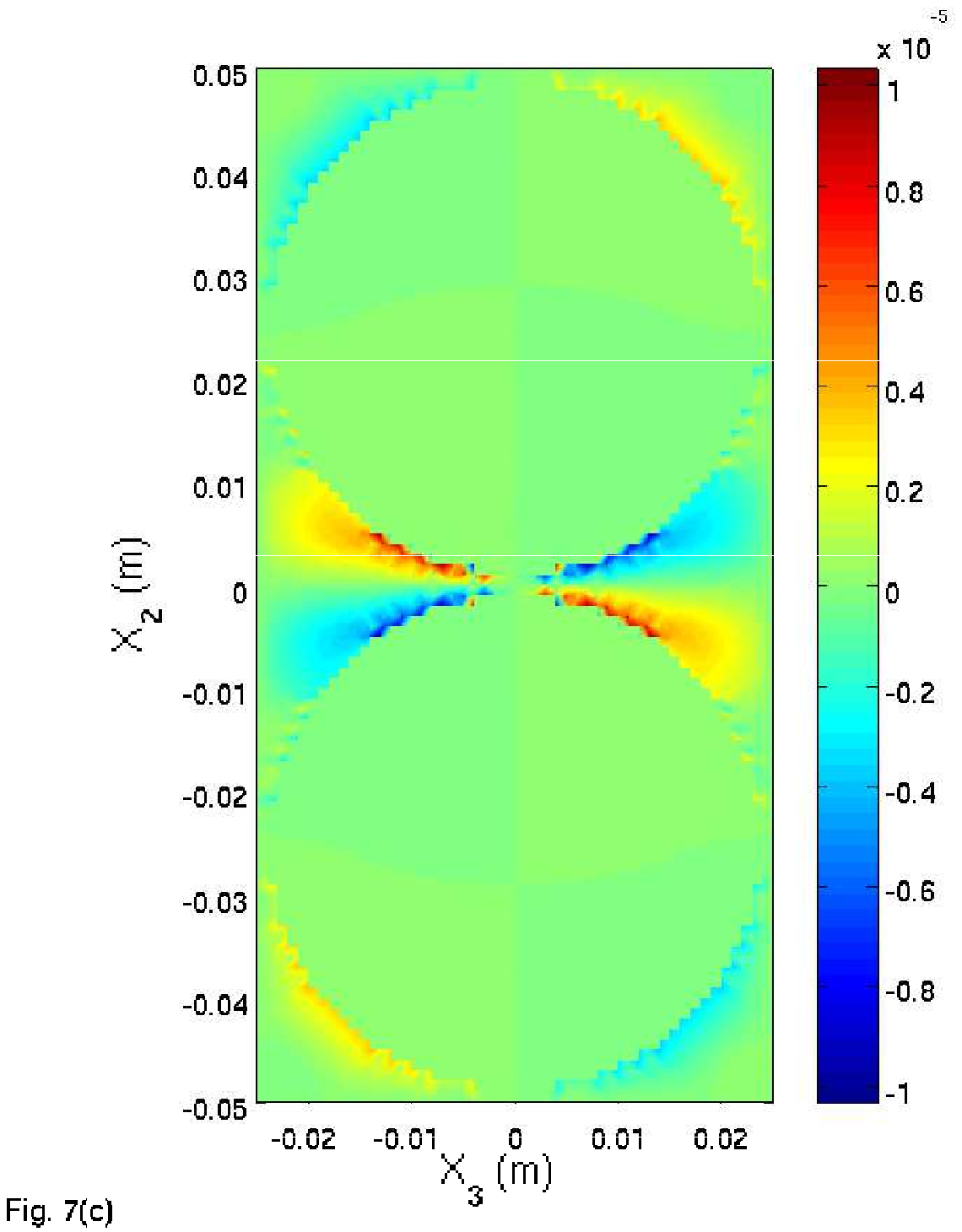}}
{\includegraphics[scale = 0.6]{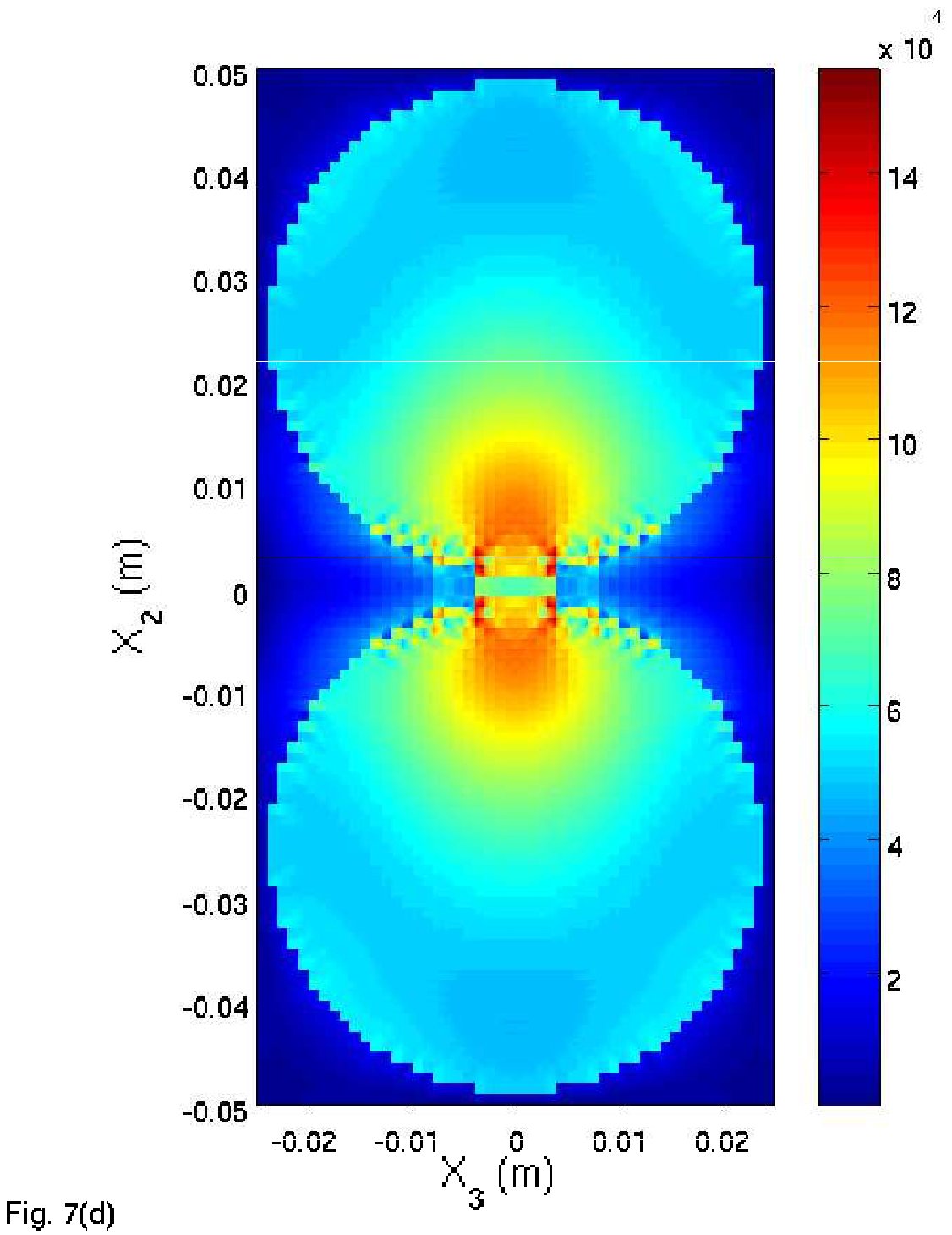}}}
\end{center}
\caption{\small The distribution of the strain components and equivalent (von Mises) stress in the iron/epoxy composite ($v_f$ = 0.7) subjected to $E_0$ = 0.003 V/m, with Nu = 1000. 
                  (a) $\epsilon_{22}$, (b) $\epsilon_{33}$, (c) $\epsilon_{23}$, and (d) $\sigma_{eq}$ (Pa). }
\label{Fig7}
\end{figure}

\noindent the ponderomotive force, which causes transverse fiber-contraction and intra-fiber attraction
that lead to compressive strain, and, in the same time heating, causing thermal expansion. 

In Fig. \ref{Fig6}(a), concerning $\epsilon_{22}$ for Nu = 1, although one might have expected maximum compressive strain in the narrow piece
of epoxy locked between the two fibers, the strain is actually mostly expansive there.
The reason is the fact that the temperature increase there is as high as it gets within a fiber, whereas the ponderomotive force vanishes there,
since it is in the axis of symmetry where it changes sign. Overall compressive strain does exists.
It is vertically around the centers of the fibers, closer to the global boundary, where the ponderomotive force is maximal,
there is little epoxy to be thermally expanded, but enough to be influenced by compressive stresses from the iron.
Fig. \ref{Fig6}(b) shows the distribution of $\epsilon_{22}$, which does not appear counterintuitive.
The maximum expansion is at the boundaries around the level of the centers of the fibers.
The ponderomotive force there causes only displacement toward the center of the specimen, without strain.
Thermal expansion therefor dominates. Fig. \ref{Fig6}(c) shows $\epsilon_{23}$, which appears to have maximum values at directions positioned forty-five degrees
to the principle axes of the specimen, as expected in linear isotropic thermo-elasticity.
Fig. \ref{Fig6}(d) shows the second-invariant (von Mises) equivalent of the deviatoric stress in each subcell.
Due to multiplication of strain by elastic moduli, it is, as expected, a simple map, showing high stress at the stiff iron fibers and lower stress in the epoxy.
The stresses ratio is an order of magnitude, though the moduli ratio is two orders of magnitude.
This owes to the fact that the strains in the epoxy are an order of magnitude higher, which in turn results from thermal domination in the strains,
along with an order of magnitude higher thermal expansion coefficient in the epoxy.

Fig. \ref{Fig7} shows the mechanical response for $E_0$ = 0.003 V/m for the case of forced convection with Nu = 1000. This case is representative of the purely magnetic effect on strain.
The temperature increase is so small, due to efficient heat transfer, that nearly the entire specimen is in compression in the direction of $x_2$, as shown in Fig. \ref{Fig7}(a).
Perhaps the most interesting feature of the entire example is the short stripe of epoxy just between the two fibers, where the compressive strain $\epsilon_{22}$ is maximal.
The advantage of Fig. \ref{Fig7}(a) is that it is representative of the macroscopic physics in the Nu$\gg$1 case:
two current-bearing wires embedded in insulation, attracted to each other, thus compressing the epoxy between them.
Fig. \ref{Fig7}(b) shows $\epsilon_{33}$, in which the prominent feature is the expansive strain in the epoxy circumventing the fibers in
the region between them, where the ponderomotive force is small in absolute value and the coefficient of thermal expansion is large.
Fig. \ref{Fig7}(c) shows the typical picture of extremal shear stresses in forty-five degrees to the symmetry axes, only unlike in Fig. \ref{Fig6}(c),
the larger strains are in the middle of the specimen, since the strains in the edges are more cylindrical, given that the ponderomotive force
is compressive in both directions and the thermal strain is an order of magnitude smaller and cannot change the sign of the ratio of the 22 and 33 strain components near the specimen edges.
Fig. \ref{Fig7}(d) is very interesting although intuitive. It shows distinctive stress concentration in the fibers around the area of their contact
with the layer of epoxy that separates them. This is what one would expect in the case of purely attractive forces,
which is effectively the situation when there is efficient forced convection removing all current-generated heat.
One notes the nice round isostress contourlines of atmospheric magnitude in the proximity of the near-contact region.

Figs. \ref{Fig8} and \ref{Fig9} show overall specimen horizontal and vertical strain-response curves for varying applied electric field, varied in a range retaining the constitutive assumptions reliable.
Fig. \ref{Fig8} shows the normal stresses in both directions for increasing $E_0$ values for Nu = 1, after averaging the strain over the specimen volume
(which is justified, strain being volume integrable in the small-strain limit).
The quadratic nature of the response is preserved, the ponderomotive force being quadratic in current and the equilibrium equation being linear in strain.
Both volume-averaged strain components are positive, signaling that the principle effect for Nu = 1 is thermal expansion.
One notes that $\epsilon_{33}$ is somewhat larger than $\epsilon_{22}$, which should be attributed to the fact that in the direction of $x_2$ there is more iron
to produce overall larger ponderomotive force. As can be seen from Fig. \ref{Fig3}, $\underset{x_2,x_3} {\text{max}} \ \bar{F}_2(x_2,x_3)\approx 2 \ \underset{x_2,x_3} {\text{max}} \ \bar{F}_3(x_2,x_3)$.
Consequently, the negative part of $\epsilon_{22}$ is larger than in $\epsilon_{33}$, and thus it is smaller by a finite observable amount.

\begin{figure}[H]
\begin{center}
{\includegraphics[scale = 0.75]{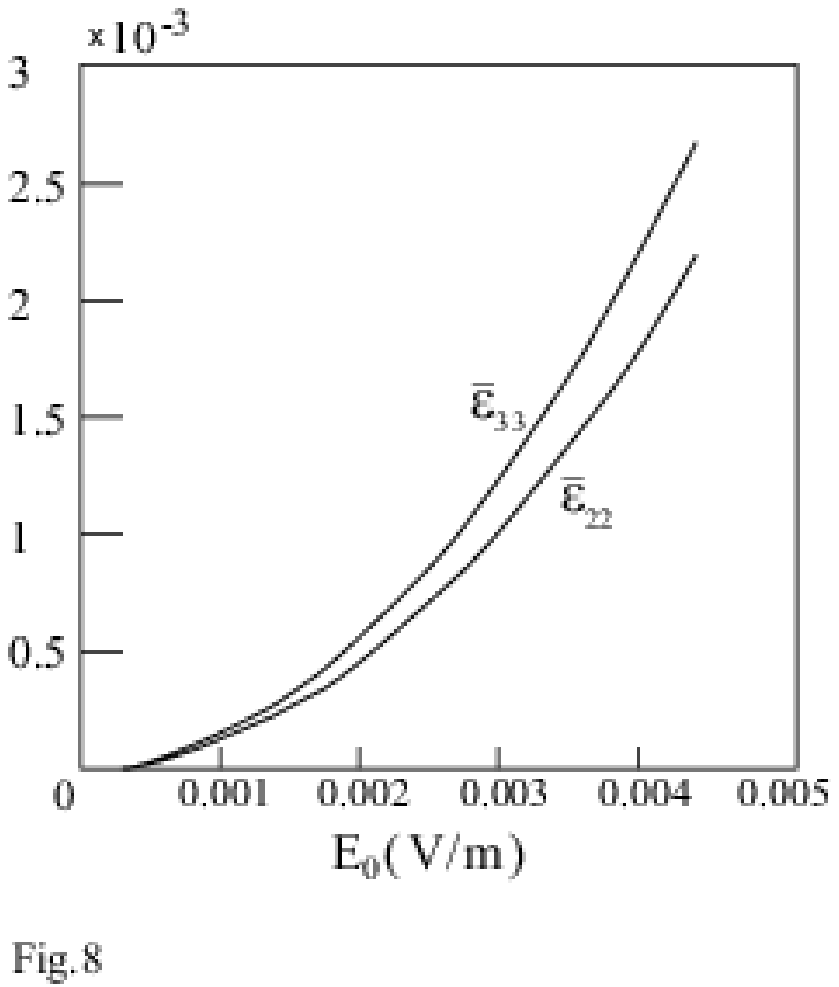}}
\end{center}
\caption{\small The variations of average strain components $\bar \epsilon_{22}$ and $\bar \epsilon_{33}$ with the applied electric field $E_0$ in the iron/epoxy composite 
                  with $v_f$ = 0.7 and Nu = 1.}
\label{Fig8}
\end{figure}

Fig. \ref{Fig8} is interesting from the second perspective of the present work, which is computational micromechanical analysis
of the steady-state response of a conductive-fiber--insulative-matrix composite  to electric field,
in the numerical scale of electrical-engineering applications (that is, where an insulated large-scale cable with two wires is
considered -- even though usually not both wires in a cable bear current), using the strong-form approach to the solution of the multi-physics boundary value problem. 

Fig. \ref{Fig9} serves the two additional perspectives of this research.
One is the construction and illustration of a computational apparatus supporting a sensing/actuating device.
Due to the large Nusselt number, thermal expansion is negligible and thus the average strain response of the specimen for Nu = 1000 shown in Fig. \ref{Fig9} is negative,
that is, there is macroscopic compression, as one would expect from a two-wire insulated cable.
The reason why this result is more illustrative of the idea of an actuator is also the fact that both average compression and expansion can be obtained
if the currents in the wires are of equal or opposite signs. Although it is assumed throughout the paper that the electric field is given,
the same apparatus works if instead the current in each wire is given. Then, by applying opposite sign currents, expansion $\epsilon_{22}$ values
can be obtained, as well as compression for equal sign currents. This, together with the fact that air cooling can be applied easier in the case of a device
than just in a general electric cable application, makes the result in Fig. \ref{Fig9} be representative of the third perspective of the paper as stated in the introduction, i.e, that of device engineering (sensor/actuator).

Finally, the dashed curve in Fig. \ref{Fig9} represents the first perspective of the paper as stated in the introduction, which is the examination of the quantitative
effect of accounting for the magnetization part of the ponderomotive force in the mechanical response in the case of a ferromagnetic conductor (in the linear regime).

\begin{figure}[H]
\begin{center}
{\includegraphics[scale = 0.7]{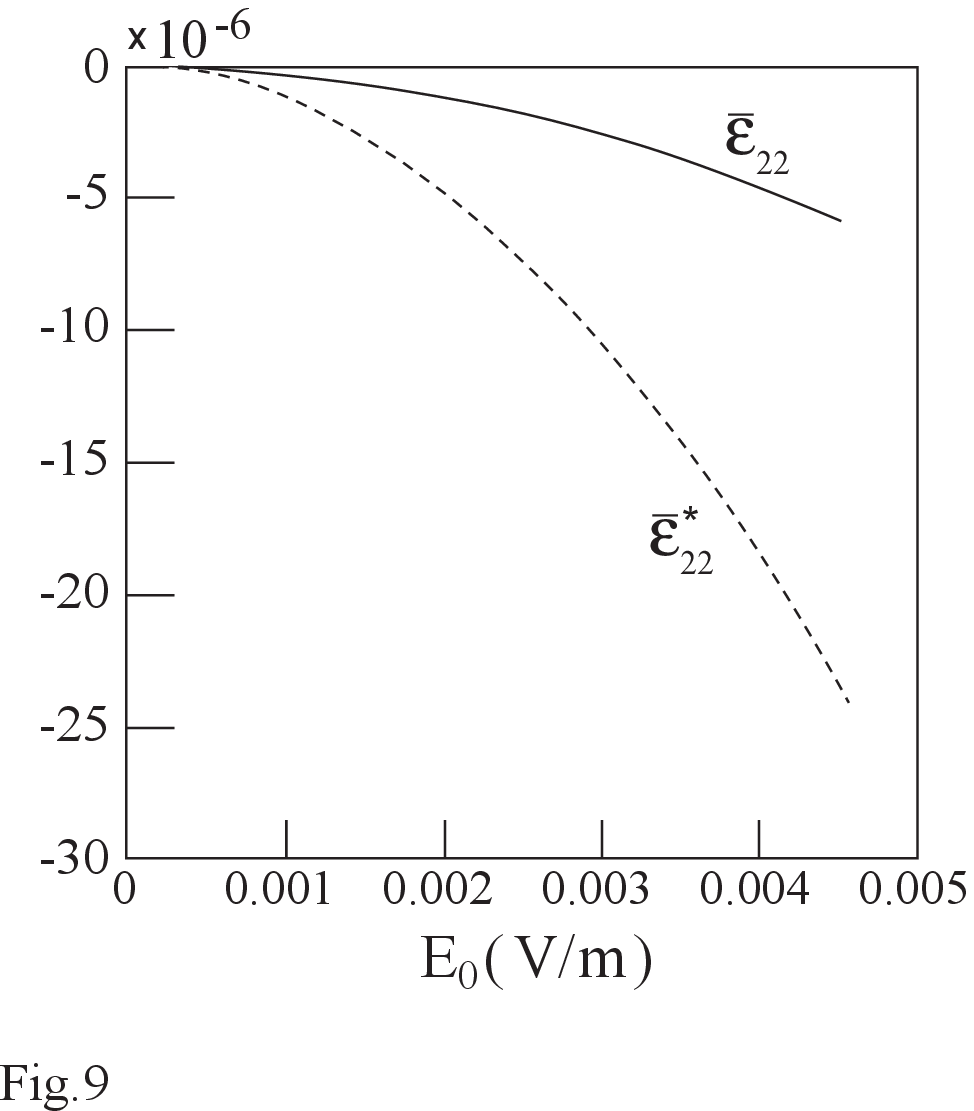}}
\end{center}
\caption{\small The variation of the volume average of strain component $\bar \epsilon_{22}$ and that of $\bar \epsilon^{*}_{22}$ with the applied electric field $E_0$
                  in the iron/epoxy composite with $v_f$ = 0.7 and Nu = 1000.    
                  The solid and dashed lines show, respectively, the strain when the forces given by either Eq. (\ref{S6}) or (\ref{A3}), are employed in the analysis.}
\label{Fig9}
\end{figure}

Clearly, neglecting magnetization results in overestimation of the average compression strain of the specimen.
This is rather intuitive. The surprising part is that although, as shown in Fig. \ref{Fig4}, the total ponderomotive force is about half the Lorentz force part,
Fig. \ref{Fig9} shows that the average strain when neglecting the magnetization force part of the ponderomotive force, is not twice larger than the actual strain,
but about five times larger (in absolute value). The reason for this apparent discrepancy is in the interplay with thermal expansion,
which is still present in the iron fibers even for Nu = 1000. In the case of calculation with the total ponderomotive force, the magnetic part of $\epsilon_{22}$ in the
fibers nearly cancels-out the thermal strain there. The result is that $\epsilon_{22}$ in the fibers can add nearly nothing to the volume-averaged value.
If the magnetization part of the force is neglected, the magnetic strain in the fibers overcomes the thermal expansion,
and the total negative-strain contribution from the fibers adds to the average value.
This effect is stronger than doubling, since the nearly zero strain exists in the realistic calculation not only in the fibers but almost in the entire specimen
except for in a small layer between the fibers. Finally, the strain in the aforementioned layer is also at least doubled when only the 'particle-in-vacuum' Lorentz force is considered.   

\section{Conclusions}
\label{Sect9}

In this article, a theory was presented, in which a computational procedure for micromechanical steady state response analysis of
a thermally conducting linear elastic composite with electrically conductive ferromagnetic fibers embedded in an insulating matrix, was derived.
The crux of the work is the construction of a stable computational scheme for the derivation of the ponderomotive force induced by an applied electric field.
The expression for the force, along with thermal analysis is consequently used in conjunction with mechanical analysis.
The mechanical analysis discretizes the domain into subcells, every one of which is filled with homogeneous material.
The equations of elasticity are solved for the array of subcells in strong form -- that is not by minimizing a variation of a functional,
as in the finite element approach, but by directly solving the boundary value problem. Quadratic expansion of 
the displacement vector within each subcell is employed. This approach, although producing a stable solution scheme for the elastic problem,
does not allow solving the magnetic problem, where quadratic or any other standard interpolation produces a singular matrix relation between
the interpolation coefficients and the edge-averaged values. Due to this fact, a different approach was taken here,
in which linear combination of point-wise analytical solutions of the differential operator representing the magnetic problem was employed, consequently
establishing the coefficients by enforcing local boundary conditions in the integral sense. 
It should be noted that in the case of the magnetic problem, additional benefit was gained from constructing the solution from point-wise analytical
solutions of the differential equations. This gain is embodied in the fact that the introduction of an auxiliary gauge condition,
such as the Coulomb gauge, became unnecessary, since the magnetic field was not assumed to be the curl of an auxiliary vector field,
but was rather constructed as the superposition of the Amp\`ere solution in a conductor and the gradient of a harmonic function. This is, of course,
only possible in the linear case.

In the thermal problem, an approach similar to the one taken in the magnetic case was chosen, although quadratic interpolation and fulfillment of the
differential equation only in the integral sense could have been an alternative possible route.

With the ponderomotive force obtained analytically for each subcell, and the temperature field obtained analytically up to four
coefficients for each subcell determined numerically, the mechanical problem was solved consistently, enforcing mechanical equilibrium on
the subcell level and requiring continuity of the integral quantities of surface forces on subcell boundaries, as well as spatially integrated displacements.

The numerical example solved with the derived theory showed reasonably intuitive results, which in itself is a certain validation of the method.
It was demonstrated that the approach is suitable for the multi-physics analysis of a current-bearing two-wire cable,
where ponderomotive and thermal effects interplay is captured quite reasonably. Furthermore,
it is the assertion of this paper that the derived theory can be used as the computational apparatus for, and in fact suggests
a sensing/actuating device with quadratic mechanical response to either uniform electric field, if a straight long wire is considered,
or the large-scale gradient of electric field, if a two-wire thin ringed cable is considered.
The present analysis is limited to steady-state response of such a device. 
Future work may extend the analysis to include transient response. 

Regarding the response to applied electric field, it may also be noted that the ability to respond to the large-scale gradient of electric field,
rather than to its local value, which may be obtained in a thin ringed device (owing to the fact that current
will be induced in the ring only in a nonuniform electric field -- the local smoothness is required, though, for the electric problem
to be solved by an electric field uniform throughout the composite's cross-section), may be useful in sensing/applications where high sensing resolution is required.

{   }

\end{document}